\documentclass{revtex4}
\usepackage{graphicx}
\begin{document}
\bibliographystyle{revtex}


\title{Particle Astrophysics and Cosmology: Cosmic Laboratories
for New Physics (Summary of the Snowmass 2001 P4 Working Group)}



\author{Daniel S. Akerib}
\email[]{akerib@phantom.PHYS.cwru.edu}
\affiliation{Department of Physics, Case Western Reserve
University, Cleveland, OH 44106}

\author{Sean M. Carroll}
\email[]{carroll@theory.uchicago.edu}
\affiliation{Department of Physics and Enrico Fermi Institute, 
University of Chicago,
5640 South Ellis Avenue, Chicago, IL 60637}

\author{Marc Kamionkowski}
\email[]{kamion@tapir.caltech.edu}
\affiliation{California Institute of Technology, Mail Code
130-33, Pasadena, CA 91125}

\author{Steven Ritz}
\email[]{ritz@milkyway.gsfc.nasa.gov}
\affiliation{NASA/Goddard Space Flight Center, Mail Code 661, Greenbelt, MD 20771}


\date{\today}

\begin{abstract}
The past few years have seen dramatic breakthroughs and
spectacular and puzzling discoveries in astrophysics and
cosmology.  In many cases, the new observations can only be
explained with the introduction of new fundamental
physics.  Here we summarize some of these recent advances.  We
then describe several problem in astrophysics and cosmology,
ripe for major advances, whose resolution will likely require
new physics.
\end{abstract}

\maketitle

\long\def\comment#1{}

\section{Overview}
The goal of the Snowmass 2001
P4 working group was to identify opportunities for
advances at the interface of
particle physics, astrophysics, and cosmology.
Since the
previous Snowmass meeting (Snowmass96 \cite{sm96}), there have been spectacular
advances in cosmology and particle astrophysics.  These include,
but are not limited to, cosmic microwave background (CMB)
evidence favoring inflation, supernova and CMB evidence for
negative-pressure dark energy, and results from solar- and
atmospheric-neutrino experiments.  Taken together, these results
demonstrate that astro/cosmo/particle physics is an integral
component of the particle-physics research enterprise.

The P4 Working Group covered a very broad range of topics,
subdivided into eight topical groups:
\begin{enumerate}
\item Dark matter and relic particles
\item Gamma rays and X-rays
\item Cosmic microwave background and inflation
\item Structure formation and cosmological parameters
\item Cosmic rays
\item Gravitational radiation
\item Neutrino astrophysics
\item Early Universe and tests of fundamental physics
\end{enumerate}
There is intense experimental and theoretical activity in all of
these areas.  Every major breakthrough listed below has been
made since Snowmass96, and the prospects for the next
decade and beyond are even brighter.   It is not possible to
describe here all the work discussed during the P4 sessions at Snowmass.
Instead, we emphasize the overarching themes that these
areas represent and their relevance to the purpose of
Snowmass 2001.

There have been spectacular observational breakthroughs:  

\begin{itemize}
\item Recent CMB measurements provide evidence that
the total energy density of the Universe, $\Omega_{\rm tot}$, is
close to unity.  For the first time,
we may know the geometry of the Universe.  Observations support
the hypothesis that large-scale structure grew from primordial
density fluctuations, in agreement with predictions from
inflation.  This provides the scientific connection between the
large-scale structure of the Universe and elementary particle
physics, and is indicative of new physics at higher energy
scales. 

\item The discrepancy between a matter density
$\Omega_m \simeq 0.3$ and $\Omega_{\rm tot} \simeq 1$
provides independent corroboration of the remarkable recent
supernova-survey evidence for some form of ``dark energy''.  If
confirmed, this suggests that $\sim70\%$ of the energy density of the
Universe is of a previously unknown and mysterious type.  The
existence of the dark energy was not even suspected by most
physicists at the time of Snowmass96.  

\item The CMB data verify that 25\% of the density of the
Universe must be in the form of nonbaryonic dark
matter---as suggested earlier by dynamical measurements of the
matter density and big-bang-nucleosynthesis predictions of the
baryon density---implying physics beyond the standard model.  This
strengthens the case for some form of particle dark matter
(e.g., supersymmetric particles or axions) in our Galactic
halo.

\item At the same time that the case for
nonbaryonic dark matter continues to strengthen, prospects
for detecting dark-matter particles over the next decade, using both
direct and indirect methods, are promising.  A fleet
of experiments with complementary sensitivities and systematics
are probing deeper into theoretically well-motivated regimes of
particle-physics parameter space.

\item Underground observations of solar neutrinos and
cosmic-ray-induced atmospheric neutrinos, indicating the
existence of neutrino oscillations, have provided evidence that
neutrinos are not massless.  This is the first direct
experimental confirmation that the standard model is
incomplete.

\item The most massive black holes are central to
TeV-class astrophysical accelerator systems that have been
observed to radiate immense power in gamma rays.  Unprecedented
leaps in sensitivity will be made by the next generation of both
ground-based and space-based gamma-ray instruments, opening up
large discovery spaces.  Together, these gamma-ray measurements
also provide a unique probe of the era of galaxy formation, and
will provide the first significant information about the
high-energy behavior of gamma-ray bursts.

\item The evidence for the highest-energy cosmic-ray
events ($E>10^{20}$ eV) poses significant challenges to our
theoretical understanding.  Observations under way, and new
detectors under construction and in planning, both on the ground
and in space, will shed new light on this highest-energy
mystery.  These experiments, as well as high-energy neutrino
telescopes, will allow us to exploit the highest-energy
particles for particle physics. 

\item Soon, the Universe will be viewed not only with
photons, but also with high-energy charged particles,
high-energy neutrinos, and gravitational waves.  These observations
will test strong-field general relativity and open vast new
windows on the highest-energy astrophysical phenomena and the
early Universe.
\end{itemize}

Interest in astro/cosmo/particle physics has grown remarkably
since Snowass96.   Of the five P working
groups at Snowmass 2001, the subscription to P4 was second only
to that of P1, Electroweak Symmetry Breaking, and the overlap
with all of HEP was obvious in P4 joint sessions with other
groups.  Interest in the Snowmass-wide teach-in on
astro/cosmo/particle physics was very strong.

In summary, there is vigorous and fast-growing activity in
astro/cosmo/particle physics, and the intellectual overlap
between the particle-physics community and others---especially
the high-energy astrophysics and cosmology communities---has
grown into full and healthy partnerships, greatly accelerating
progress.  These partnerships provide enormous opportunities as 
well as new challenges.

\section{Introduction}

The aim of particle physics is to understand the fundamental
laws of nature.  The primary tools in particle
physics have been, and continue to be,
accelerator experiments, as they provide
controlled environments for addressing precise
questions.  However, such experiments have limitations,
especially when we consider that many of the most promising ideas
for new particle physics---e.g., grand unification and quantum
gravity---can be tested only at energies many
orders of magnitude greater than those accessible by
current and planned accelerators.

During the past few decades, it has become increasingly apparent 
that the study of cosmology and astrophysics can provide hints
or constraints to new physics beyond the standard model.  In
just the past few years, the promise of these endeavors has
begun to be realized experimentally, most notably with new
results from the CMB, evidence for an accelerated cosmological
expansion, and with evidence of neutrino masses and mixing that
comes from experiments with neutrinos from astrophysical
sources.

For Snowmass 2001, we convened a working group
aimed to discuss how
cosmology and astrophysics can be used in the future to search
for new physics beyond the standard model.  We organized our
activities into eight topical subgroups, each convened by three
to six sub-group convenors.  These groups and the convenors were:
\begin{itemize}
\item  Dark matter and relic particles
(S. Asztalos, P. Gondolo, W. Kinney, and R. Schnee)

\item  Gamma rays and X-rays (J. Buckley,
T. Burnett, and G. Sinnis)

\item  The CMB and inflation (S. Church, A. Jaffe,
and L. Knox)

\item  Structure formation and cosmological
parameters (R. Caldwell, S. Deustua, P. Garnavich, L. Hui,
T. McKay, and A. Refregier)

\item  Cosmic rays (J. Beatty, J. Mitchell,
P. Sokolsky, and S. Swordy)

\item  Gravitational radiation (P. Bender,
C. Hogan, S. Hughes, and S. Marka)

\item  Neutrino astrophysics (B. Balantekin,
S. Barwick, J. Engel, G. Fuller, and T. Haines)

\item  The early Universe and tests of fundamental
physics (A. Albrecht, J. Frieman, and M. Trodden)
\end{itemize}
P4 working group plenary talks and parallel
sessions took place throughout all three weeks of the workshop, 
often in coordination with the activities of the E6 working
group.  The P4 sessions attracted the participation of well over
200 scientists from particle physics and astrophysics.

In this article, we briefly summarize the activities of the P4
working group with an aim of listing science questions for
particle physics that may be addressed with astrophysical and
cosmological experiments and observations.  More detailed
discussions of all of these topics, as well as more
comprehensive lists of references, will be provided by the
individual topical sub-group reports.

\section{Dark Matter and Relic Particles}

Since the mystery of dark matter first appeared in the thirties due to
observations of galaxy clusters by Zwicky, the evidence has steadily
mounted and today strongly suggests the possibility of a solution
rooted in new fundamental particle physics~\cite{jungman}.  
Dark matter refers to matter that is inferred only through its
gravitational effects, and which neither emits nor absorbs
electromagnetic radiation. These effects are observed on a wide
range of distance
scales---from individual galaxies to superclusters to mass flows on
the largest observable scales. Direct observation of dark matter and
the determination of its nature is one of the most important
challenges to be met in cosmology today. Moreover, it is likely that
this determination will yield new information in particle physics,
since there is strong evidence that the dark matter is not composed of
baryons, but rather is in some exotic form.

At the time of Snowmass96, much of the evidence for
nonbaryonic dark matter was already in place.  The most reliable
measurements of the universal mass density
come from galaxy clusters, the largest observed
equilibrium structures.  Using a variety of methods (gravitational
lensing, virial analysis, X-rays from intracluster gas), it is found
that the matter density is significantly greater than can be accounted
for by the baryon density allowed by big-bang nucleosynthesis (BBN)
and the measured primordial abundance of deuterium and other light
elements (e.g., \cite{copi,burles,nollett}). As discussed
elsewhere in this document,
cosmological evidence accumulated since Snowmass96 provides
independent confirmation of this picture.  Observations of distant
supernovae and the cosmic background radiation together suggest a flat
Universe in which 30\% of the energy density is due to
nonrelativistic matter and only 4\% due to baryons, consistent
with the measurements from clusters and BBN, respectively.

\begin{figure}
\includegraphics[width=.8\textwidth]{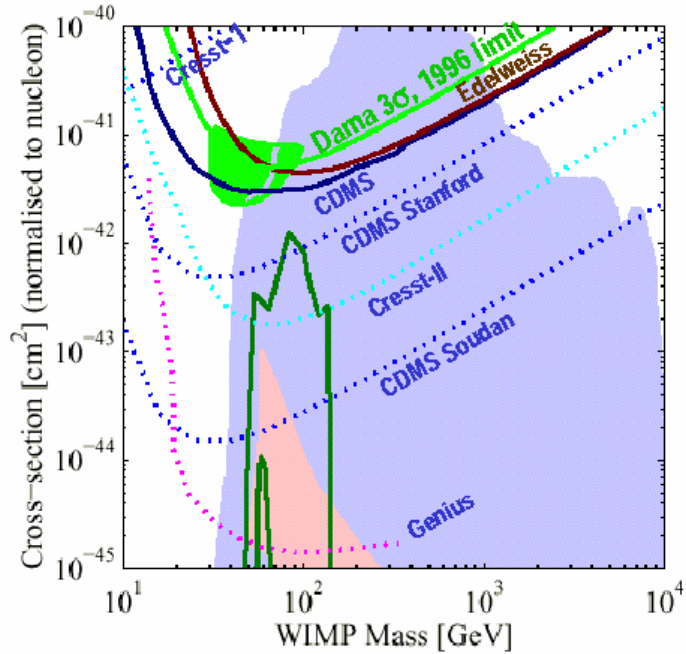}
\caption{\label{fig-direct} Regions of the WIMP
mass--cross-section parameter space accessible with various
detectors. The solid lines indicate upper limits (90\% C.L.) 
achieved to date, along
with the 3-sigma contour for the DAMA experiment. The CDMS limit is 
the dark solid line. The projected limits for future experiments 
are shown as
dashed lines. The large shaded areas show regions of parameter space
allowed by different supersymmetric models; see
{\tt http://dmtools.berkeley.edu} for
further details and a complete list of references.}
\end{figure}

Particle physics offers two different hypotheses for the dark 
matter---WIMPs and axions---either
of which would constitute a major discovery of physics beyond the
standard model.  First, weakly-interactive massive particles, or
WIMPs, would be produced in thermal equilibrium with the hot plasma of
the early Universe. WIMPs with masses in the 10--1000~GeV/$c^2$
range and
weak-scale cross sections would have fallen out of equilibrium in
sufficient number to have a relic density today comparable to the
critical density. The leading and best studied candidate in this
generic class of particles is the lightest superpartner (LSP) from
supersymmetry, which is expected to be the
neutralino~\cite{jungman}. Experimental approaches 
to detecting dark matter were discussed in
great detail in the E6 sessions.  There are direct laboratory
searches for nuclear recoils produced by WIMP-nuclear
interactions; searches for energetic neutrinos from
annihilation of WIMPs that have accumulated in the Sun and/or
Earth; and searches for exotic cosmic-ray antiprotons and
positrons and gamma rays produced by WIMP annihilation in the
Galactic halo.  We note here that a broad range of
experiments, including many new technological developments, are
steadily improving the sensitivity for detecting WIMPs in the galactic
halo, both through direct WIMP-nuclear scattering experiments and
through indirect searches. Current searches are already
exploring the parameter space of supersymmetric WIMPs, with
prospects for a factor of a hundred improvement in the coming
five years~\cite{cdms,dama,edelweiss}.

The two most sensitive experiments to date appear to be
inconsistent. The DAMA collaboration \cite{dama}
measures an annual modulation in
the event rate in a large array of NaI scintillators, which is
expected for WIMPs due to the differential speed of the Earth about
the galactic center as it orbits the sun.  However, the results are
controversial within the dark-matter community, since the possibility
of a systematic effect has not been ruled out. Also, the CDMS
experiment \cite{cdms}, using a smaller array of more sensitive
lower-background detectors, does not observe the expected event 
rate that should occur if the WIMP inferred by DAMA has scalar
interactions with nuclei, while null searches for energetic
neutrinos from the Sun conflicts with DAMA if the WIMP has
spin-dependent interactions with nuclei \cite{Ullio:2000bv}.
The signal and limits from these
experiments are shown in Figure~\ref{fig-direct} (for scalar interactions), 
along with a number of previous and proposed experiments.  While
it is clear that a positive result would be a landmark discovery in
both astrophysics and particle physics, it is also clear that
complementary studies in the more controlled setting of
accelerator-based experiments will be crucial in unraveling the
relevant physics.

In addition to a review of progress in the field of direct and indirect
WIMP detection, there were cross-cutting sessions devoted to the
connections between halo WIMP searches and SUSY searches at
accelerators~\cite{ellis}, as well as current ideas of galaxy formation
that predict steep dark-matter spikes around the black hole at
the Galactic center~\cite{GS,ullio}. These spikes lead to
predictions for higher-than-observed rates of gamma-ray annihilation
products.  There were also discussions about possible
discrepancies between observations and theoretical models that
predict halo cusps \cite{NFW} and substructure \cite{Moore,Klypin}
that have lead some researchers to consider other
particle dark-matter candidates, such as warm or
self-interacting dark matter~\cite{spergel}.
These were very active sessions and reflect the exciting ongoing work at
the interplay between these areas of research. In particular, the
resolution of the cuspy-halo and spike questions are quite
likely to tell us something about particle physics and galaxy
formation~\cite{GS,ullio}.

\begin{figure}
\includegraphics[width=.8\textwidth]{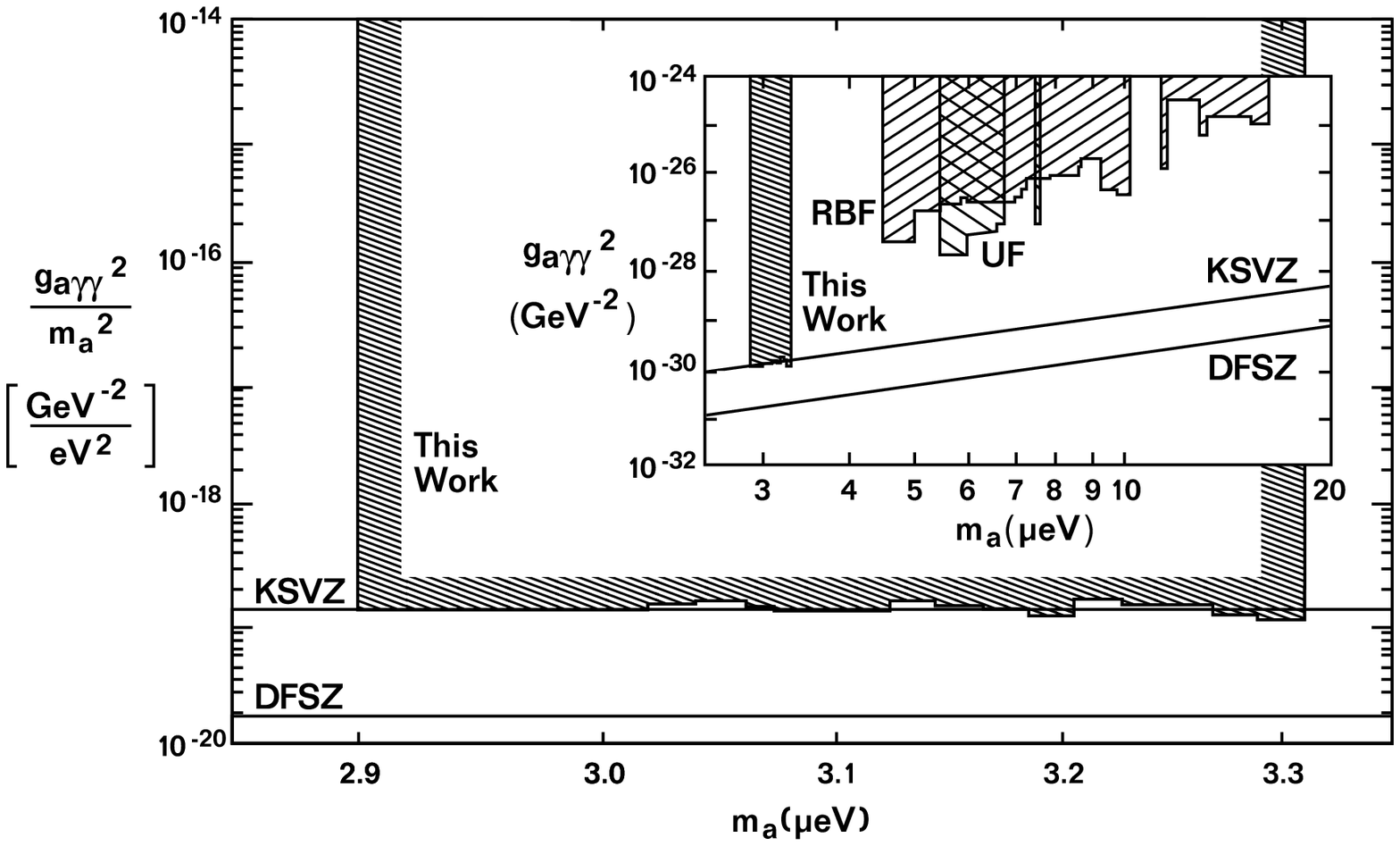}
\caption{\label{fig-axion} Regions of axion mass-coupling
parameter space currently being probed by an ongoing search at
Livermore \cite{Asztalos}.}
\end{figure}

The second important candidate for particle dark matter is the axion,
the pseudo-Nambu-Goldstone boson resulting
from the spontaneous breakdown of Peccei-Quinn
symmetry, a hypothetical symmetry which has been 
postulated to explain the lack of CP violation in the strong
interaction~\cite{rosenberg}.  Axions would be produced as a Bose
condensate during the QCD phase transition, or through the decay of
axionic strings or domain walls. The best method conceived to detect
axions is to exploit their pseudoscalar coupling to two photons. By
threading a high-Q RF cavity with a strong magnetic field, axions
would interact with the B-field photons and produce final-state
photons with energy nearly equal to the axion mass. The allowed mass
range for the axion can be studied by sweeping the resonant frequency
of the cavity. When the cavity is tuned to the frequency that matches
the axion mass, the cavity will resonate with excess power. The
experimental challenge of sufficiently low cavity/amplifier noise has
been met and realistic ``KSVZ'' axion models corresponding to $2.9 <
m_a < 3.3~\mu eV$ have now been tested and ruled out
(Fig.~\ref{fig-axion}).  Work is ongoing
to extend the mass range and lower the noise floor so that nearly two
decades of mass can be probed at the weaker coupling of the ``DFSZ''
model.  The two decades of axion mass that have not yet
been ruled out by experiments or astrophysical observations are
precisely in the range that could explain the dark matter. It is
reasonable to expect that in less than a decade, axions as dark matter
could be detected or definitively ruled out.  Moreover, the RF cavity
experiments arguably stand the best chance of any approach for
discovering axions at all.

\section{Gamma-ray astrophysics}

\subsection{Introduction}

Measurements of the fluxes of celestial gamma rays provide
a variety of important information about the Universe.
Gamma rays are emitted by particle jets from nature's
largest accelerators and they otherwise do not
interact much at their source, offering a direct view inside.
Similarly, the Universe is mainly transparent to gamma rays below
10 GeV, so they probe cosmological volumes, and the
energy-dependent attenuation of the flux above 10 GeV provides
important information about those volumes.  Conversely, gamma rays
interact readily in detectors, with a clear signature; and they
are neutral, so there are no complications due to magnetic fields
(galactic-flux calculations do not have trapping-time
uncertainties, the photons point directly back to their sources,
etc.).  In general, gamma-ray emission identifies sites of
extreme particle acceleration and/or decays of very massive states.
The gamma-ray energy regime is one of the least well-measured
portions of the electromagnetic spectrum, and the measurements
that have already been done have confirmed the expectation of
surprise when a new domain is opened for exploration.  At
Snowmass, we reviewed recent results, surveyed the next
generation of experiments and their capabilities, and explored
the physics opportunities provided by the upcoming experiments.
Sessions included talks and discussions on areas of shared
interest with other sub-fields, notably dark matter, X-rays,
cosmic rays, and neutrinos.

\subsection{Experiments}

\begin{figure}
\includegraphics[width=0.8\textwidth]{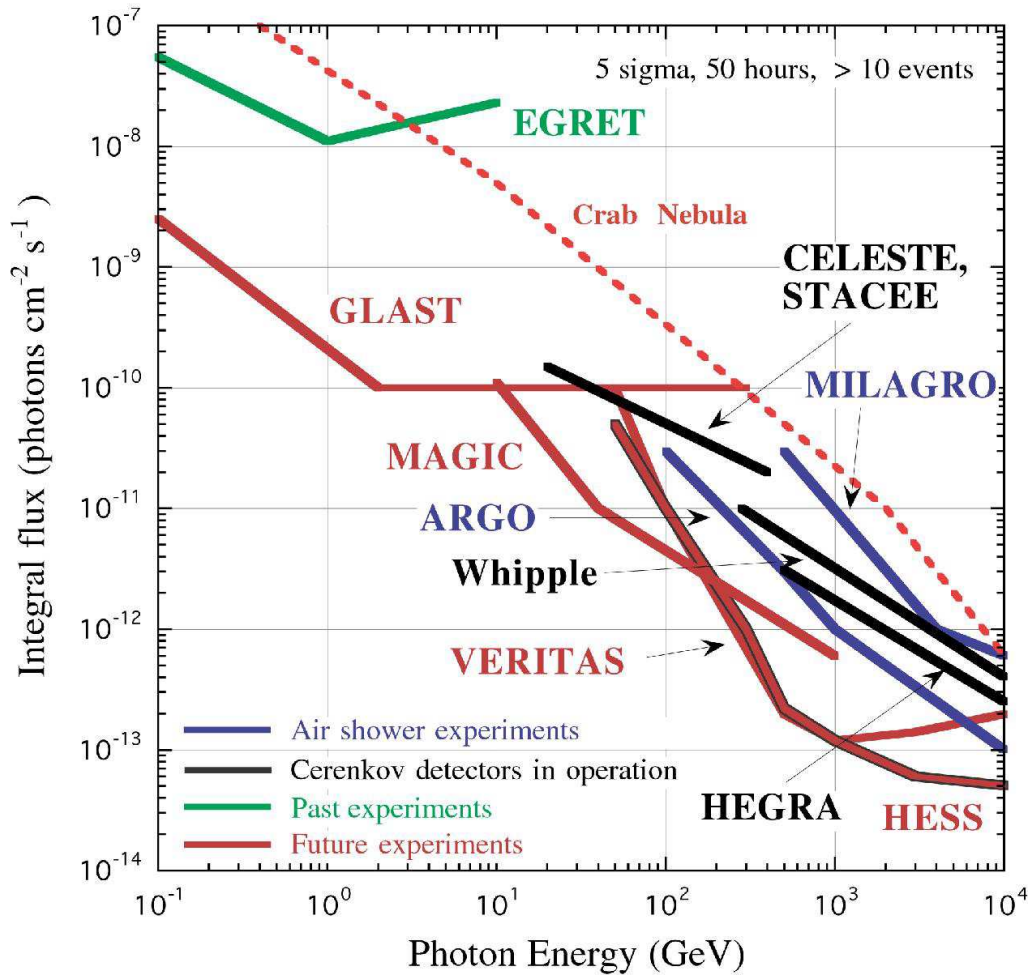}
\caption{\label{fig-gamma_experiments} Sensitivities and
frequency ranges for current and planned gamma-ray experiments
\cite{morselli}.}
\end{figure}

Although gamma rays with energies below 10 GeV have no difficulty
reaching us from the edge of the visible Universe, they interact
in the upper atmosphere and never reach the ground.
Space-based measurements are required.  For significantly
higher-energy gamma rays, sufficient information from the air showers
initiated by the primary photon reaches the ground and can be
measured with ground-based apparatus.  Space-based detectors
use the pair-conversion technique for direction information and
separation from the large cosmic-ray backgrounds.  There are two
basic types of ground-based detectors: air-shower Cerenkov
telescope (ACT) detectors, and air-shower particle detectors.
Since the flux of celestial gamma rays is falling rapidly with
increasing energy (typically as $E^{-2}$ or faster), larger
collecting area is required for meaningful statistics at higher
energy. Large-area space-based detectors are a challenge. One of
the main goals of the present and next round of experiments,
therefore, is to push the high-energy frontier of space-based
instruments upward and the low-energy threshold of ground-based
instruments downward, to provide a significant overlap in
coverage.  A one-decade or more energy overlap will provide
important opportunities for systematic error comparisons and
essential information about both the gamma-ray sources themselves
and the energy-dependent flux attenuation in this particularly
interesting region.  The importance of future experiments with
sensitivity comparable to that of VERITAS for $E>100$ GeV,
but with a much wider field of view (FOV), was also discussed at
Snowmass.

The experimental situation is summarized in Figure
\ref{fig-gamma_experiments}.  The vertical axis is the
point-source sensitivity integrating over photons with energy
greater than $E_0$, given on the horizontal axis. The two
spaced-based experiments are toward the upper
left, and the ground-based experiments are toward the lower
right.  The ground-based experiments have a much better
sensitivity in photons cm$^{-2}$s$^{-1}$, but astrophysics
dictates that they must.  The red dashed line gives the Crab flux
integrated above $E_0$, showing the rapid decrease in flux with
increasing energy.  The sensitivity to physical sources is, in
some sense, therefore given by the perpendicular distance from
the flux line.  Viewed in that way, the sensitivities of the
ground-based and space-based experiments in each generation are
quite well matched.

Ground-based and space-based detectors have capabilities that are
complementary.  Ground-based air Cerenkov detectors typically have
good angular resolution, low duty cycle, huge collecting area,
small FOV, and good energy resolution.
Ground-based air-shower particle detectors have relatively poor
energy resolution, but much larger duty cycle and FOV.
Space-based detectors have good angular resolution, excellent
duty cycle, relatively small collecting area, excellent FOV, and
good energy resolution with relatively small systematic
uncertainties.  Since the gamma-ray sky is extremely variable,
and spans more than seven orders of magnitude in energy, it is
important to operate these detectors together.

Gamma-ray experiments draw the interest of both particle
physicists and high-energy astrophysicists.  An important feature
of this field is that, instead of having separate collaborations
of particle physicists and astrophysicists that periodically
share their results, the communities are already intertwined
within the collaborations.  The designs of the instruments and
the effectiveness of the collaborations have benefitted greatly
from this deepening cooperation.

 \subsection{Abridged overview of physics topics}

\begin{figure}
\includegraphics[width=0.6\textwidth]{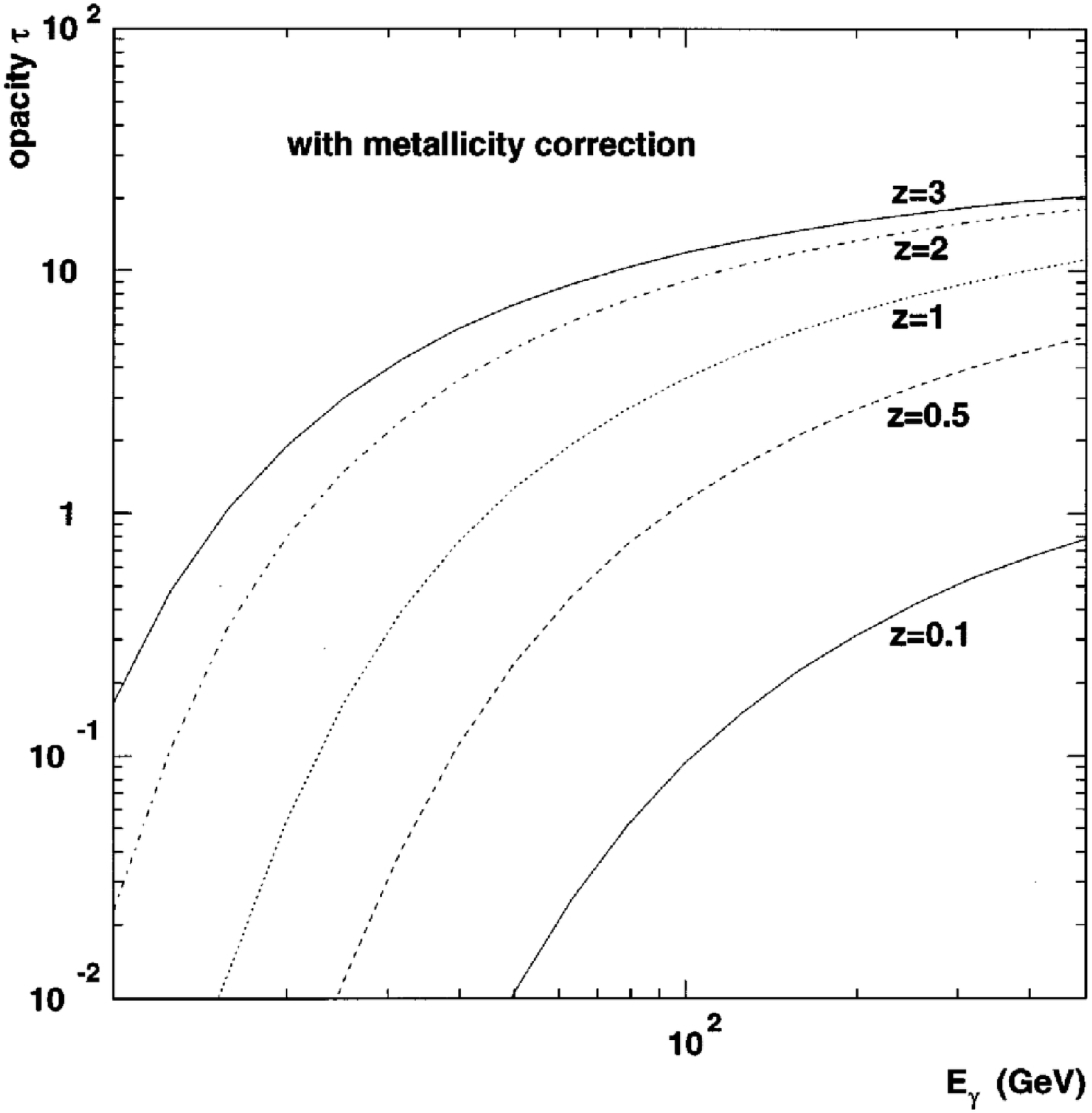}
\caption{\label{fig-EBL} Opacity for the extragalactic
background light as a function of photon energy for curves of
constant redshift~\cite{Stecker}.}
\end{figure}

The flux of $E>10$ GeV photons is attenuated due to pair
creation by the diffuse field of UV-optical-IR extragalactic
background light (EBL).  The EBL is produced
mainly by starburst activity, so a measurement of the EBL gives
important information about the era of galaxy formation.
Different models \cite{Stecker,Primack,madau} predict distinctly
different EBL densities.  Figure \ref{fig-EBL} shows the opacity
as a function of energy for curves of constant redshift
for one model.  Using
known gamma-ray point sources, the TeV gamma-ray experiments
study the EBL in a relatively local volume of space, while the
1--100 GeV experiments study the EBL density over cosmological
distances.  This is a major motivation for intermediate
experiments such as STACEE \cite{STACEE} and CELESTE \cite{CELESTE}.

The gamma-ray sky includes an apparently diffuse, isotropic flux
($\approx 1.5\times 10^{-5} {\rm cm}^{-2}~{\rm s}^{-1}~{\rm
sr}^{-1}$ for $E>100$
MeV \cite{Hunter}) that is presumably extragalactic.  The
origin of this flux is a mystery.  Is it really isotropic,
produced at an early epoch in intergalactic space, or an
integrated flux from a large number of yet-unresolved point
sources?  GLAST
\cite{GLAST1,GLAST2}, with a large effective area (for a space
experiment) and much-improved angular resolution, will have the
capability to resolve a large fraction of the flux, if it is
indeed composed of point sources. This constitutes a `no lose'
theorem: either the diffuse flux will be resolved into thousands
of new point sources (compared with the total catalog of 271
sources from EGRET \cite{EGRET}) 
to be studied both in detail in their own right and for
use in EBL probes, or a truly diffuse flux of gamma rays
from the early Universe will have been discovered.

There is an excellent candidate for those point sources: active
galactic nuclei (AGN), which produce vast amounts of power (some
flares are believed to reach $10^{45}$ ergs/sec)
from a very compact volume.  The prevailing idea 
\cite{blandfordrees,blandfordkonigl,rees84}
is that AGN are powered by accretion onto supermassive black
holes ($10^6 - 10^{10}$ solar masses).  Models of AGN include
multi-TeV, highly collimated particle jets with extremely
variable gamma-ray emission. Due to the huge collecting area, the
next-generation ACTs such as VERITAS 
\cite{VERITAS} will have the remarkable
capability to resolve sub-15-minute variability of these objects,
providing important information about the characteristic sizes of
the gamma-ray emitting regions.  Constraining models of AGN
high-energy emission requires observations across all
wavelengths, so once again use of the the combined capabilities
of all the experiments is essential.

Gamma-ray bursts (GRBs) are the most powerful known explosions in
the Universe, and their origin continues to be a mystery.  While
the behavior of bursts at lower energies is now being studied in
greater detail, even less is known about the behavior of bursts
at the highest energies.  This situation should improve
dramatically over the coming decade: if the bursts are close
enough, or bright enough, their cutoff energies may finally be
observed by the ACTs.  The air-shower experiments such as Milagro
\cite{MILAGRO} and ARGO \cite{ARGO}, with their large FOV and 
excellent duty cycle, may
serve as alert triggers.  There is already evidence that
Milagrito, the Milagro prototype, has observed TeV emission from
one burst \cite{Atkins}.
The GLAST observatory should see
hundreds of bursts, covering the energy range 10 keV to 300 GeV
and also providing worldwide burst alerts.

Bursts at cosmological distances may also allow novel searches
for a small velocity dispersion of photons, since the vast
distance provides a huge lever arm.  Models 
\cite{Amelino-Camelia:1998gz} that
anticipate a full theory of quantum gravity suggest such
dispersion.  Due to the broad energy coverage, these measurements
can be made within a single gamma-ray experiment.  The signature
would be an arrival-time ordering of events with energy.
However, the intrinsic emission characteristics of bursts must be
understood.  The arrival-time differences should increase with
redshift for a true dispersion, providing an additional
observational handle.  Evolution effects would then have to be
ruled out if such a signal were observed.

If the galactic dark matter is a halo of WIMPs, annihilations of
these particles could produce a detectable flux of gamma
rays.  This possibility is especially intriguing since one
signature would be mono-energetic `lines', set by the WIMP mass,
at somewhere between 10's and 100's of GeV.  One of the open
questions is the distribution of WIMPs, particularly near the
galactic center, since the annihilation rate is proportional to
the square of the density.  The degree to which the density
spikes has a critical impact on other exclusion limits and on the
size of the potential gamma-ray signal \cite{GS,ullio}.

Additional topics include pulsars, supernova remnants and the
origin of the cosmic rays (at least for $E<10^{14}\, {\rm eV}$), the
unidentified sources from EGRET, emission from galactic clusters,
primordial-black-hole evaporation, decays of topological
defects and other massive relics from the big bang, and
signatures of large extra dimensions \cite{Arkani-Hamed:1998nn}.

Finally, we note that this list
only contains the phenomena we already know, or that we have
already imagined.  It is reasonable to expect that the
next decade of astrophysical gamma-ray observations will be
punctuated with surprising discoveries.

\section{The Cosmic Microwave Background and Inflation}

\subsection{Recent Progress in the CMB}

In forthcoming years, the cosmic microwave background (CMB) will 
provide one of the most exciting opportunities for learning
about new physics at ultra-high-energy scales (for recent
reviews, see, e.g., \cite{KamKos99,HuDod01}, as well as
the report of the P4.3 working group \cite{church}).  
The past two years have already seen spectacular advances in
measurements of temperature fluctuations in the CMB
that have led to major advances in our ability to
characterize the largest-scale structure of the Universe, the
origin of density perturbations, and the early Universe.  In the
next year we should see enormous improvements with the recently
launched MAP satellite, and then even more precise data with the
launch of Planck \cite{PLANCK} in 2007.

The primary aim of these experiments has been to determine the CMB
power spectrum, $C_\ell$, as a function of multipole moment $\ell$.
Structure-formation theories predict a series of bumps in the power
spectrum in the region $50\lesssim \ell \lesssim 1000$, arising as
consequences of oscillations in the baryon-photon fluid before CMB
photons last scatter \cite{SunZel70,PeeYu70}.  The rich structure in
these peaks allows simultaneous determination of the geometry of the
Universe \cite{KamSpeSug94}, the baryon density, Hubble constant $h$,
matter density, and cosmological constant, as well as the nature
(e.g., adiabatic, isocurvature, or topological defects) and spectrum
of primordial perturbations \cite{Junetal96}.

\begin{figure}
\includegraphics[width=0.8\textwidth]{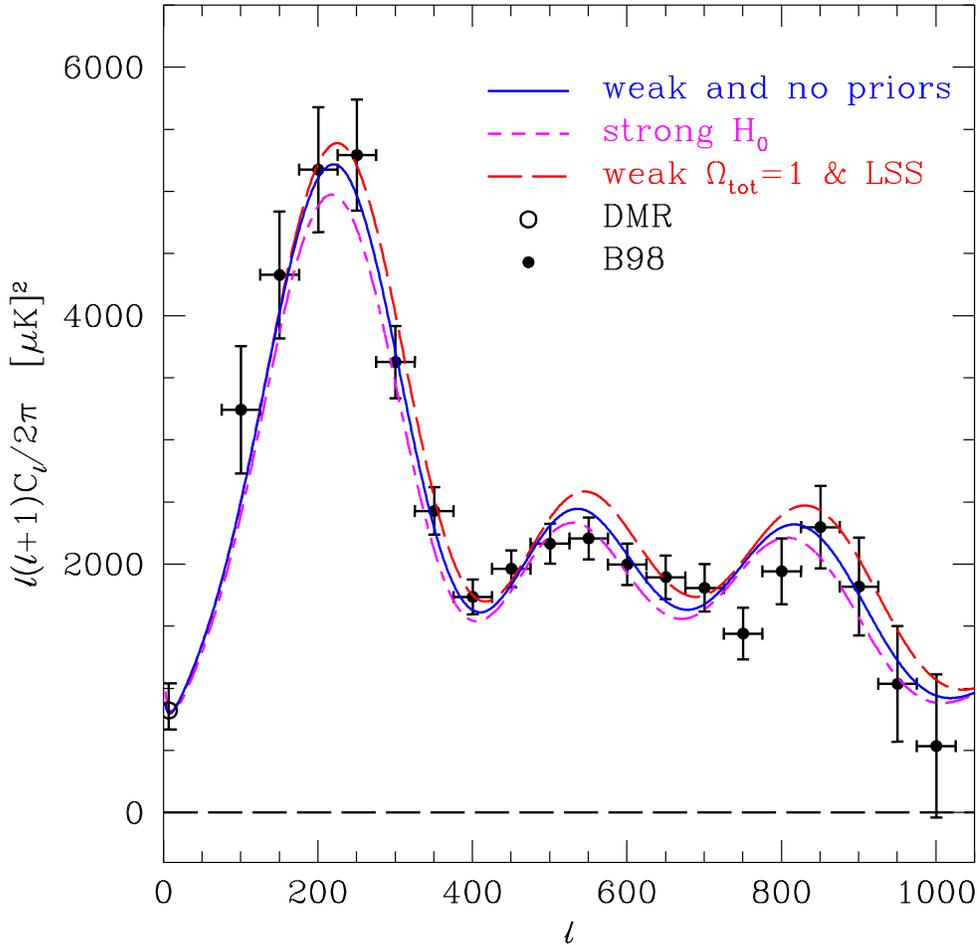}
\caption{\label{fig:boom} The CMB power spectrum measured
recently by BOOMERanG~\cite{boom}.  Similar results have been
obtained also by DASI~\cite{dasi} and MAXIMA~\cite{max}.}
\end{figure}

Within the past two years, three independent experiments that use
different techniques, observing strategies, and frequencies have each
measured the power spectrum in the range $50\lesssim\ell\lesssim1000$
with sufficient precision to see clearly a first and second peak in
the CMB power spectrum, as well as hints of a third
\cite{boom,max,dasi}, as shown in Figure~\ref{fig:boom}.  These
experiments represent a watershed event in cosmology, as they suggest for
the first time that the Universe is flat and that structure grew from
a nearly scale-invariant spectrum of primordial density perturbations.
These two properties are robust predictions of
inflation, a period of accelerated expansion in the very early
Universe driven by vacuum energy associated with
ultra-high-energy physics.

\begin{figure}
\includegraphics[width=\textwidth]{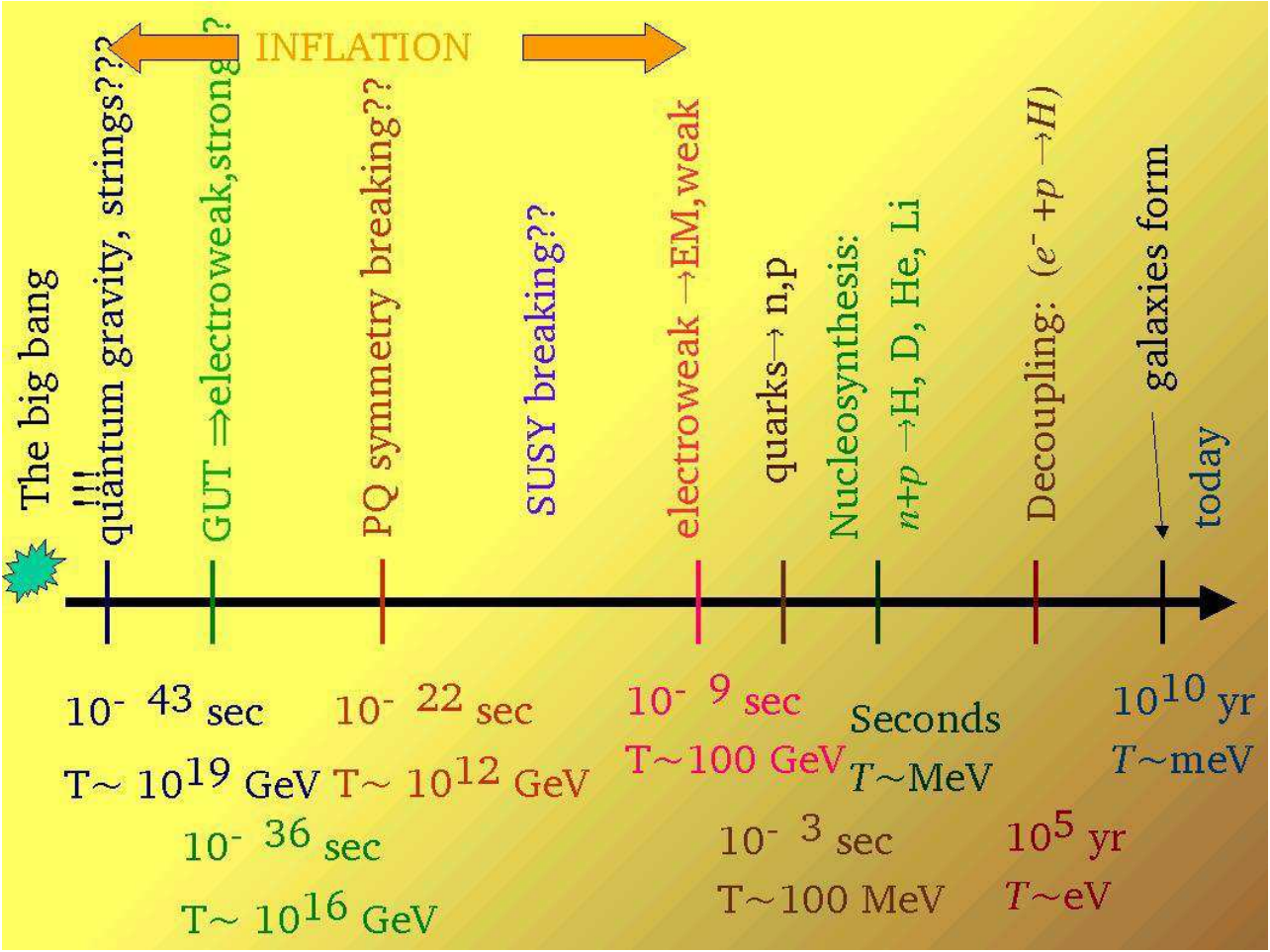}
\caption{\label{fig:timeline} Logarithmic history of the
Universe.  Inflation may have taken place at any time from the
Planck time until the time of electroweak-symmetry breaking.
Given the CMB confirmation of the inflationary predictions of a
flat Universe and primordial adiabatic perturbations, an obvious
goal of early-Universe cosmology should be to determine the
time, or equivalently, the energy scale of inflation (defined
more precisely to be the fourth root of the vacuum-energy
density during inflation).  This can
be accomplished by searching for the unique CMB polarization
pattern produced by the inflationary gravitational-wave
background, the amplitude of which scales on the inflationary 
energy scale.}
\end{figure}

Although these recent CMB tests suggest that we are on the right 
track with inflation, we still have no idea what new physics may 
have given rise to inflation.  Plausible theoretical models
place the energy scale of inflation anywhere from the
Planck scale to the electroweak scale, and associate the
inflaton (the scalar field responsible for inflation) with new
fields that arise in string theory, GUTs, the Peccei-Quinn
mechanism, supersymmetry breaking, and electroweak-scale
physics, as shown in Figure \ref{fig:timeline}.

\subsection{Inflation, Gravitational Waves, and CMB Polarization}

Perhaps the most promising avenue toward further tests of
inflation as well as determination of the energy scale of
inflation is the gravitational-wave background.
Inflation predicts that quantum fluctuations in the
spacetime metric during inflation should give rise to a
stochastic gravitational-wave background with a
nearly-scale-invariant spectrum \cite{AbbWis84}.  
Inflation moreover predicts
that the amplitude of this gravitational-wave background should
be proportional to the square of the energy scale of inflation.

These gravitational waves will produce temperature fluctuations
at large angles.  Upper limits to the amplitude of large-angle
temperature fluctuations already constrain the energy scale of
inflation to be less than $2\times10^{16}$ GeV.  However, since
density perturbations can also produce such temperature
fluctuations, temperature fluctuations cannot alone be used to
detect the gravitational-wave background.

Instead, progress can be made with the polarization of the CMB.
Both gravitational waves and density perturbations will produce
linear polarization in the CMB, and the polarization patterns
produced by each differ.  More precisely, gravitational waves
produce a polarization pattern with a distinctive curl pattern
that cannot be mimicked by density perturbations (at linear
order in perturbation theory; see below)
\cite{KamKosSte97,SelZal97}.  Moreover, inflation
robustly predicts that the amplitude of this  polarization curl
depends on the square of the energy scale of inflation.

Is this signal at all detectable?  If the energy scale 
of inflation is much below the GUT scale, then the polarization
signal will likely be too small to ever be detected.  However,
if inflation had something to do with GUTs---as many, if not
most theorists believe---then the signal is conceivably
detectable by a next-generation CMB experiment.  Although the
MAP satellite, launched just last month, is unlikely to
have sufficient sensitivity to detect the curl component from
inflationary gravitational waves, the Planck satellite, a
European Space Agency experiment to be launched in 2007, 
should have sufficient sensitivity to detect the CMB curl
component as long as the energy scale of inflation is greater
than roughly $5\times10^{15}$ GeV.  However, Planck will not be
the end of the line.  An experiment that integrates more deeply
on a smaller region of sky can improve the sensitivity to the
inflationary gravitational-wave background by almost two orders of
magnitude.  Moreover, there are several very promising ideas
being pursued now that could improve the detector sensitivity by 
more than an order of magnitude within the next decade.  Putting 
these two factors together, it becomes likely that a CMB
polarization experiment that probes inflationary energy
scales to below $10^{15}$ GeV---and thus accesses the entire
favored GUT parameter space---could be mounted on a ten-year
timescale (if not sooner).

\subsection{Cosmic shear and the CMB}

There are several interesting astrophysics questions that must be
addressed on the way to this inflationary goal.  First of all, cosmic
shear---weak gravitational lensing by density perturbations along the
line of sight---can produce a curl component in the polarization.
This cosmic-shear curl component might be subtracted (and the curl
component from inflationary gravitational waves isolated) by measuring
higher-order correlations in CMB temperature maps \cite{ZalSel99}, and
so precise CMB temperature maps must be made in tandem with the
polarization maps.  The information from these CMB cosmic-shear maps
will be of interest in their own right, as they probe the distribution
of dark matter throughout the Universe as well as the growth of
density perturbations at early times.  These goals will be important
for determining the matter power spectrum and thus for testing inflation
and constraining the inflaton potential, as discussed further in the
Section on structure formation below.

\subsection{CMB and Primordial Gaussianity}

Another prediction of inflation is that the distribution of mass 
in the primordial Universe should be a realization of a Gaussian 
random process.  This means that the distribution of temperature 
perturbations in the CMB should be Gaussian and it moreover
implies a precise relation between all of the higher-order
temperature correlation functions and the two-point correlation
function.  These relations can be tested with future precise CMB
temperature and polarization maps.

\subsection{The Sunyaev-Zeldovich Effect and Cosmological
Parameters}

A further goal of CMB studies is precise measurement of the 
Sunyaev-Zeldovich effect (see, e.g., Ref. \cite{Bir99} and
references therein), the scattering of
CMB photons from the hot gas in galaxy clusters.  Measurements
of the Sunyaev-Zeldovich effect may soon provide entirely new
and unique avenues for classical cosmological tests---e.g.,
measurements of the matter density and of the expansion history
(as discussed in connection with dark energy below), as well as
for studying the growth of large-scale structure (as discussed
in connection with inflation below).  A particularly intriguing
advantage of the Sunyaev-Zeldovich effect is that the signal
from a given cluster does not depend on the distance to the
cluster.  This allows for the possibility of detecting clusters
at much higher redshifts than may be detected with X-ray or
other observations.  Also, the amplitude of the
Sunyaev-Zeldovich effect is linear in the electron density, and
thus provides a more faithful tracer of the underlying cluster
mass than the X-rays, which depend on the more uncertain square
of the electron density.

\section{Cosmological Parameters and Structure Formation}

\subsection{Dark energy}

A long-standing aim of cosmology has been the determination 
of the contents of the Universe, characterized by their
fractional contribution $\Omega$ to the critical density.  These
components include, for example, the baryon density
$\Omega_b$ and the cold-dark-matter density $\Omega_{\rm cdm}$.
During the past few years, there has been a convergence in
measurements of these parameters from a variety of independent
techniques.  The new CMB measurements discussed above find a
baryon density $\Omega_b h^2 \simeq 0.02$, in very good
agreement with the baryon density expected from big-bang
nucleosynthesis (using the recently determined quasar deuterium
abundance \cite{burles}).  
And as discussed above, the CMB power spectrum is best fit by a dark-matter
density $\Omega_m\simeq0.3$, in agreement with a number of
dynamical determinations of the matter density.

\begin{figure}
\includegraphics[width=0.7\textwidth]{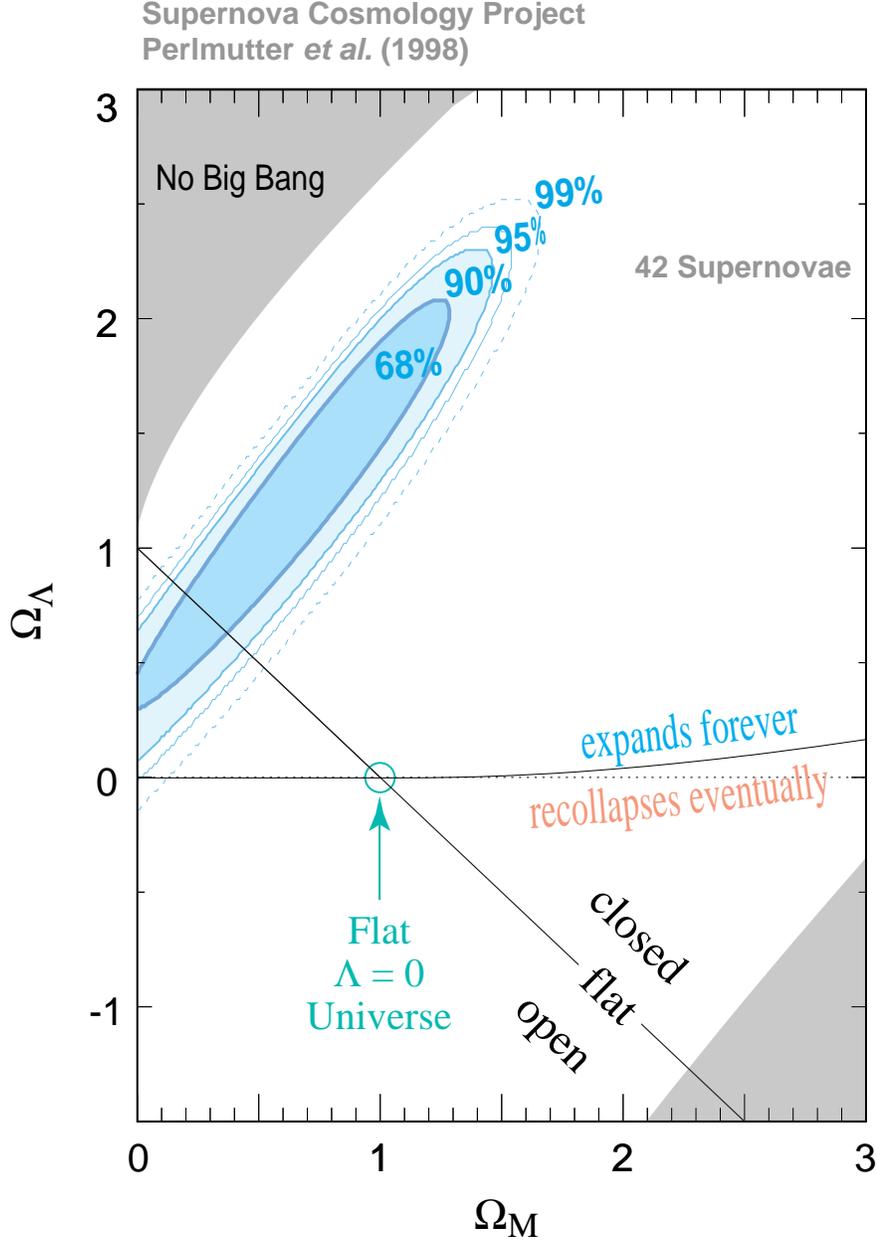}
\caption{\label{fig:snap} Regions in the
$\Omega_m-\Omega_\Lambda$ parameter space consistent with
supernova searches and the CMB (which now constrain $\Omega_m +
\Omega_\Lambda \simeq 1$).}
\end{figure}

In addition to confirming the predictions of big-bang
nucleosynthesis and the existence of dark matter, the
measurement of classical cosmological parameters has resulted in
a startling discovery over the
past few years: roughly 70\% of the energy density
of the Universe is in the form of some mysterious
negative-pressure dark energy \cite{Carroll:2000fy}.  Supernova
evidence for an accelerating Universe~\cite{Peretal99,Rieetal98}
has now been dramatically
bolstered by the discrepancy between the total cosmological
density $\Omega_{\rm tot}\simeq1$ indicated by 
the CMB and dynamical
measurements of the nonrelativistic-matter density
$\Omega_m\simeq0.3$.  New and independent evidence is provided by
higher peaks in the CMB power spectrum that also suggest
$\Omega_m\simeq0.3$, again leaving 70\% of the density of
the Universe unaccounted for.  

As momentous as these results are for
cosmology, they may be even more remarkable from the vantage point of
particle physics, as they indicate the existence of new physics
beyond the standard model plus general relativity.  Either
gravity behaves very peculiarly on the very largest scales, or
there is some form of negative-pressure ``dark energy'' that
contributes 70\% of the energy density of the Universe.
For this dark energy to accelerate the expansion, its
equation-of-state parameter ${w}\equiv p/\rho$ must satisfy
${w}<-1/3$, where $p$ and $\rho$ are the dark-energy pressure
and energy density, respectively.  The simplest guess for this dark
energy is the spatially uniform, time-independent cosmological
constant, for which ${w}=-1$. Another possibility is
quintessence~\cite{Caletal98,RatPee88,CobDodFri97,TurWhi97} or
spintessence \cite{BoyCalKam01}, a cosmic scalar field that is
displaced from the minimum of its potential.  Negative pressure is
achieved when the kinetic energy of the rolling field is less than the
potential energy, so that $-1 \le {w} < -1/3$ is possible.

Although it is the simplest possibility, a
cosmological constant with this value is difficult to reconcile
with simple heuristic arguments, as quantum gravity would
predict its value to be $10^{120}$ times the observed value, or
perhaps zero in the presence of an unknown symmetry.  
One of the appealing features of dynamical models for dark
energy is that they may be compatible with a true vacuum
energy which is precisely zero, to which the Universe will
ultimately evolve.  

The dark energy was a complete surprise and remains a
complete mystery to theorists, a stumbling block that, if
confirmed, must be
understood before a consistent unified theory can be
formulated.  This dark energy may be a direct remnant of string
theory, and if so, it provides an exciting new window to
physics at the Planck scale.

The obvious first step to understand the nature of this dark
energy is to determine whether it is a true cosmological
constant, or whether its density evolves with time.  This can be 
answered by determining the expansion rate of the Universe as a
function of redshift.  In principle this can be accomplished
with a variety of cosmological observations (e.g.,
quasar-lensing statistics, cluster abundances and properties,
the Lyman-alpha forest, galaxy and cosmic-shear surveys, etc.).
However, the current best bet for determining the expansion
history is with supernova searches, particularly those that can
reach beyond to redshifts $z\gtrsim1$.  
Here, better systematic-error reduction, better theoretical
understanding of supernovae and
evolution effects, and greater statistics, are all required.
Both ground-based (e.g., the
DMT \cite{DMT} or WFHRI \cite{WFHRI}) and space-based (e.g.,
SNAP \cite{SNAP})
supernova searches can be used to determine the expansion
history.  However, for redshifts $z\gtrsim1$, the principal optical
supernova emission (including the characteristic silicon
absorption feature) gets shifted to the infrared which is
obscured by the atmosphere.  Thus, a space-based observatory
appears to be required
to reliably measure the expansion history in the
crucial high-redshift regime.

If the dark energy is quintessence, rather than a cosmological
constant, then there may be observable consequences in the
interactions of elementary particles if they have some coupling
to the quintessence field.  In particular, if the cosmological
constant is time evolving (i.e., is quintessence), then there is 
a preferred frame in the Universe.  If elementary particles
couple weakly to the quintessence field, they may exhibit small
apparent violations of Lorentz and/or CPT symmetry (see, e.g.,
\cite{Carroll}).  A variety of accelerator and astrophysical
experiments can be done to search for such exotic signatures.

\subsection{Structure formation}

\begin{figure}
\includegraphics[width=\textwidth]{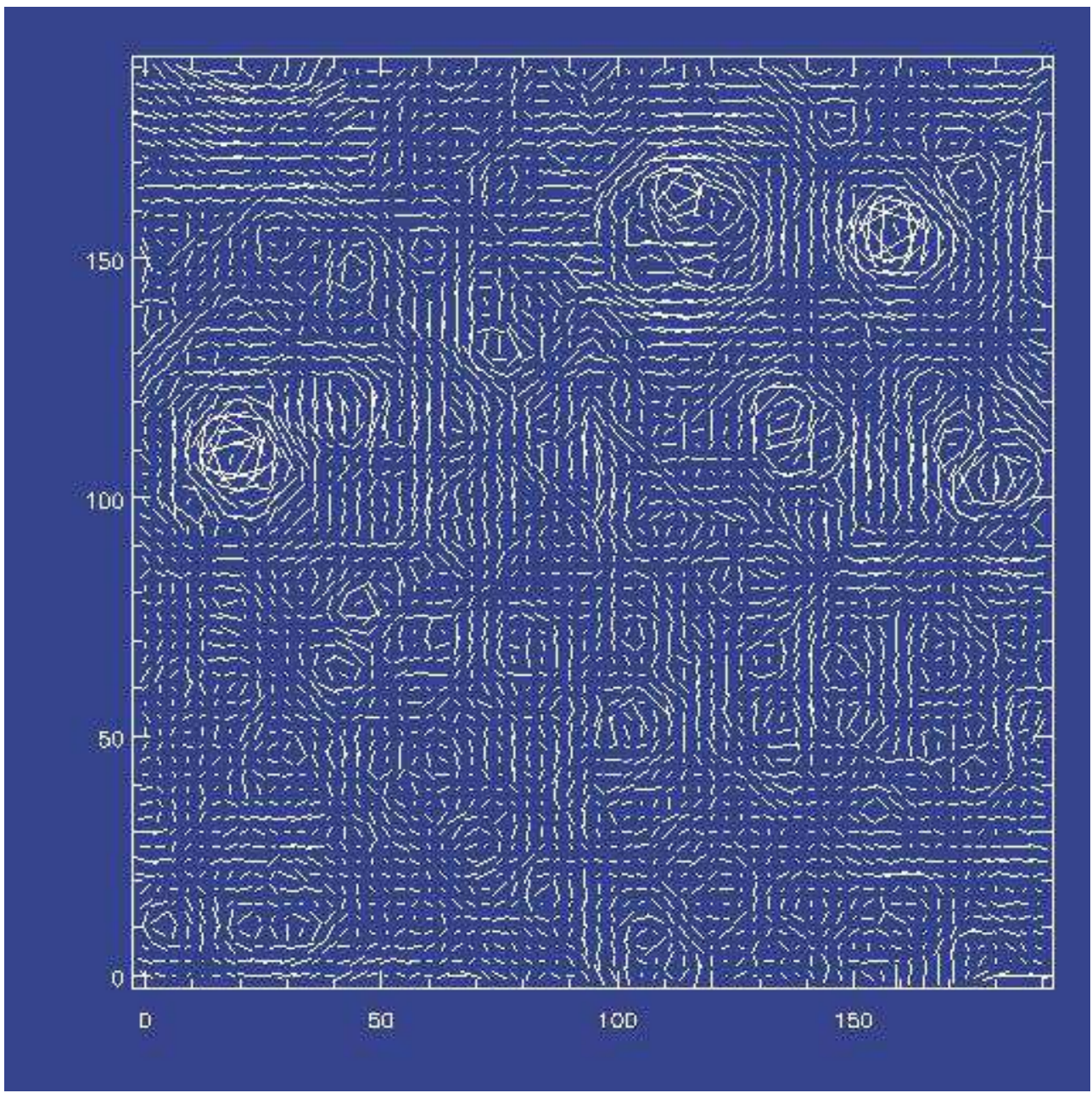}
\caption{\label{fig:shear} A simulation of a cosmic-shear
map that could be reconstructed from the alignments of the
shapes of numerous distant background galaxies.  The projected
mass density can be inferred from such a map.}
\end{figure}

Since the Snowmass96 meeting, large-scale galaxy surveys have 
become a reality, particularly with the advent of the Two-Degree
Field \cite{2dF} and Sloan Digital Sky Surveys \cite{SDSS}.  We
are now mapping the distribution
of galaxies over huge volumes in the Universe.  Moreover, just
last year, four independent groups reported detection of
``cosmic shear,'' correlations in the distortions to the shapes
of distant galaxies induced by weak gravitational lensing due to
large-scale mass inhomogeneities
\cite{KaiWilLup00,vanWetal,Witetal,Bacetal00}.
In the future,
cosmic-shear measurements will map the distribution of matter
(rather than just the luminous matter probed by galaxy surveys)
over large volumes of space  (interestingly, the
telescope requirements for cosmic-shear maps match those for
supernova searches).  Figure \ref{fig:shear} illustrates how a
cosmic-shear experiment might map the mass density projected
along the line of sight.

If the big bang is a cosmic accelerator, subtle
correlations in the debris from the explosion can provide
valuable information on inflation, just as subtle correlations
in jets in accelerator experiments can provide information about 
the collisions that give rise to them.
The primary aims of galaxy surveys and cosmic-shear maps are
determination of the
power spectrum $P(k)$ of matter in the Universe as well as
higher-order correlation functions for the mass distribution. 
These measurements are important for the study of inflation, as
inflation relates the amplitude and shape of the power spectrum
$P(k)$ to the inflaton potential $V(\phi)$ \cite{LidLyt}.
Moreover, as discussed above, inflation predicts very precise relations
between all of the higher-order correlation functions for the
primordial mass distribution and its two-point correlation
function, and these relations can also be tested with the
observed distribution of mass in the Universe today.
The growth of density perturbations via gravitational infall
alters the precise structure of the correlation hierarchy
from the primordial one.  However, it does so in a calculable
way so that the primordial distribution of density perturbations 
(Gaussian as predicted by inflation? or otherwise?) can be
determined from the distribution observed in the Universe today
\cite{threept}.

\begin{figure}
\includegraphics[width=\textwidth]{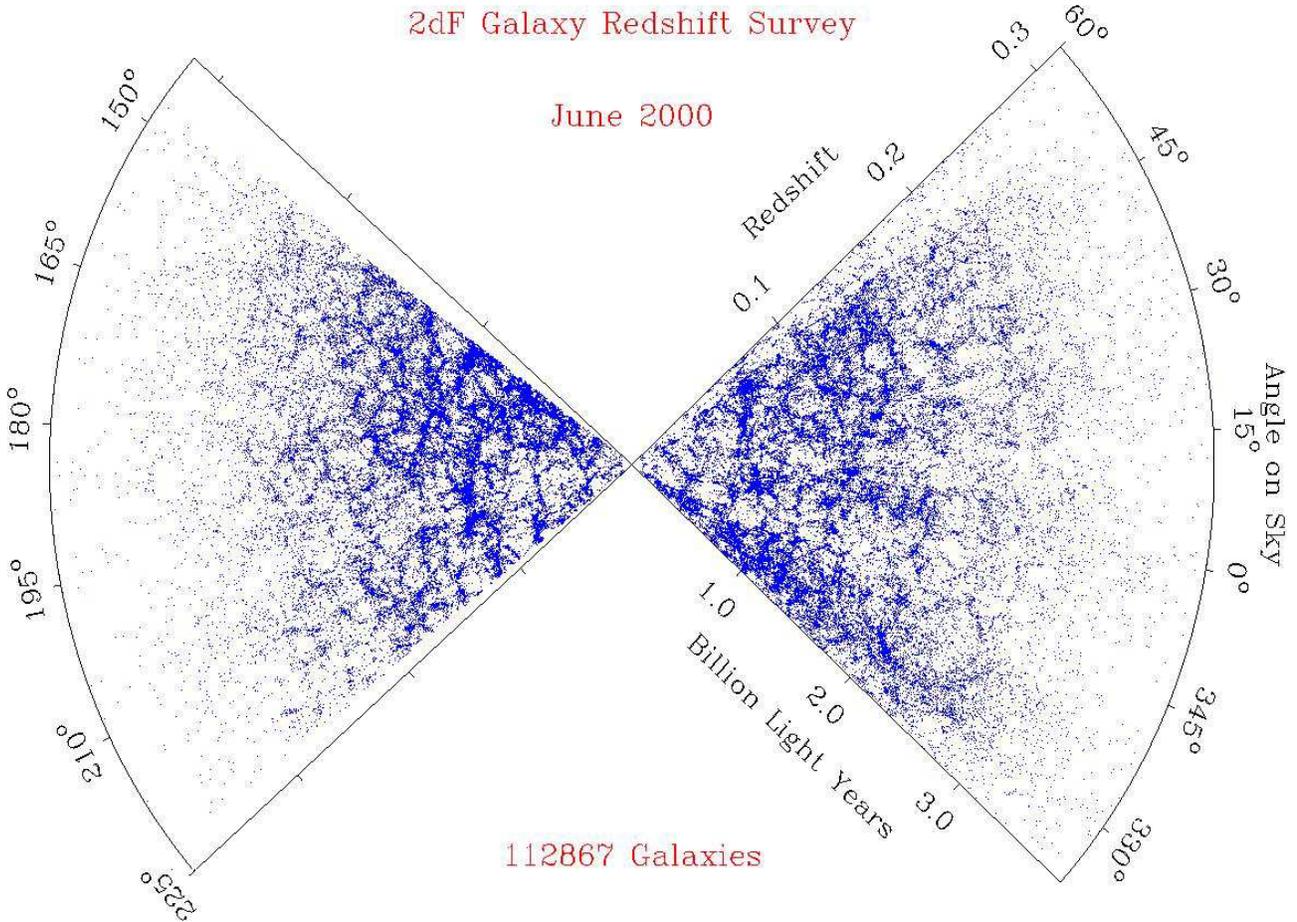}
\caption{\label{fig:2df} Galaxy distribution as mapped by
preliminary results from the 2dF survey.  According to
inflation, the galaxy distribution can be used to infer
parameters of the inflaton potential.}
\end{figure}

Information about the primordial distribution of matter can also 
be obtained by studying the abundances and properties of the
rarest objects in the Universe: clusters of galaxies today and
galaxies at high redshift (see, e.g., \cite{verde}).  Such
objects form at rare
($\gtrsim3\sigma$) high-density peaks in the primordial density
field.  Inflation predicts that the distribution of such peaks
should be Gaussian.  If the distribution is non-Gaussian---for
example, skew-positive with an excess of high-density
peaks---then the abundance of these objects can be considerably
larger.  In such skew-positive models, such objects would also
form over a much wider range of redshifts and thus exhibit a
broader range of properties (e.g., sizes, ages, luminosities,
temperatures).

\section{Cosmic Rays}

There are strong historical connections between particle physics
and cosmic-ray physics.  Over time, as artificial accelerators
reached high energies and intensities, the emphasis of cosmic-ray
research has shifted away from using cosmic rays as natural
particle beams toward their use as cosmic messengers.

\begin{figure}
\includegraphics[width=0.8\textwidth]{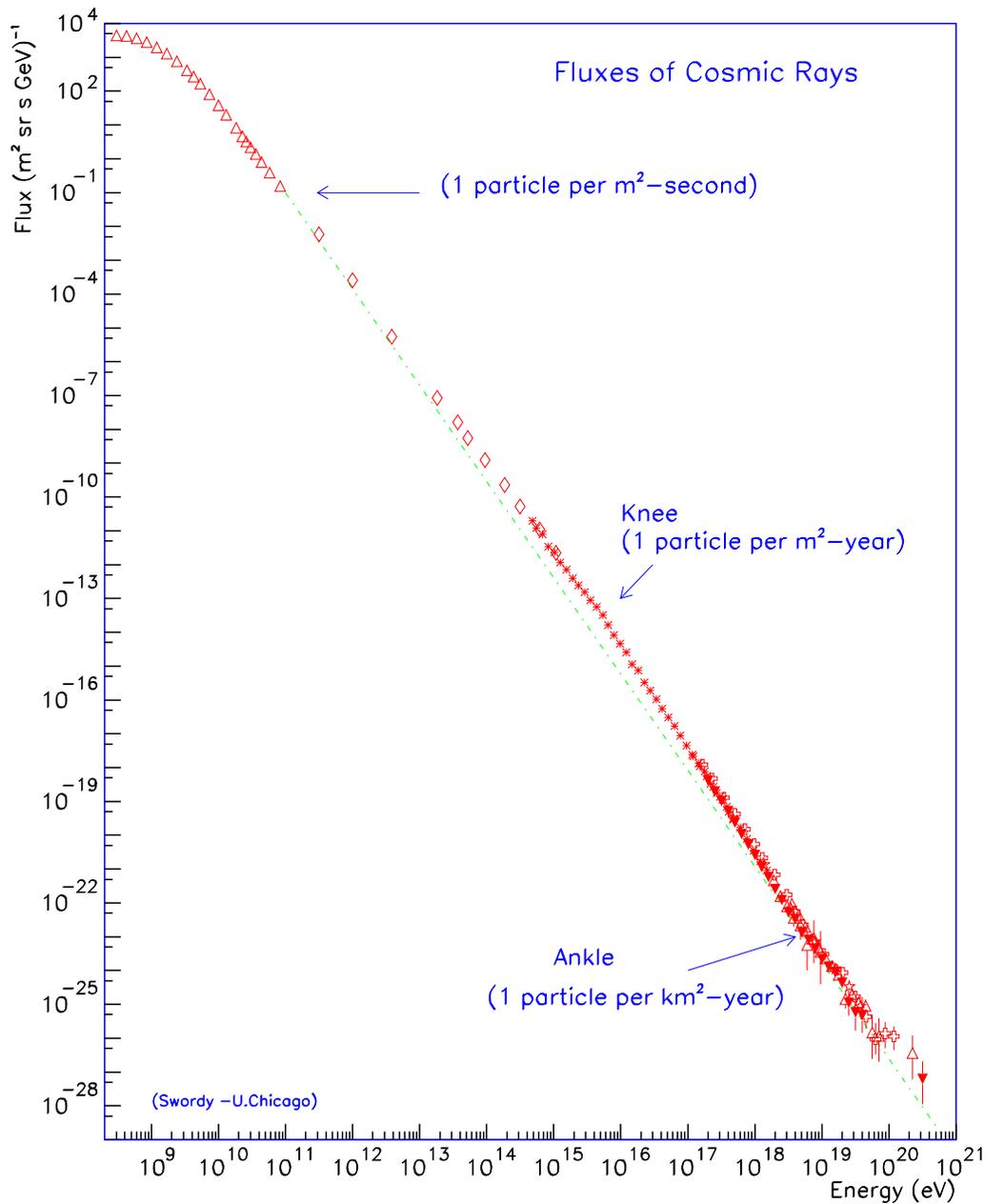}
\caption{\label{fig-CRspectrum} Cosmic-ray spectrum.}
\end{figure}

The cosmic-ray spectrum is shown in Figure \ref{fig-CRspectrum}.
Some of the questions today are (1) What is the origin of cosmic
rays?  This remains an unsolved mystery, though there are good
candidates for at least a portion of the energy spectrum.
(2) What are the processes involved in the propagation of cosmic
rays from their source(s) to us?  (3) What
produces the features seen in figure \ref{fig-CRspectrum}?  In
particular, what produces the `knee' and the `ankle'?

There are also some other questions of perhaps more direct
interest to particle physics. (4) Is there cosmic-ray antimatter
from exotic sources in the Universe?  For example, topological
defects or supermassive relics (such as those that might account
for ultra-high-energy cosmic rays; see below) might produce
particles and antiparticles in equal numbers leading to
cosmic-ray antiprotons or positrons.  Although theory says that
it is highly unlikely that large regions of the Universe consist
of antimatter, even a single anti-nucleus (e.g., anti-carbon)
would demonstrate the existence of large antimatter regions.
Finally, in some supersymmetric models, annihilation WIMPs in
the Galactic halo could produce detectable fluxes of cosmic-ray
antiprotons and/or positrons.  The AMS
experiment \cite{ref_AMS}, which had a successful engineering
flight on the shuttle in 1998, is manifested for the space
station in 2005 to help answer these questions.
The PAMELA experiment \cite{pamela1,pamela2}, which will be in
operation in early 2003, will measure the antiproton and
positron fluxes in polar orbit.

Over the past decade, balloon measurements have provided a wealth of
new information to help address questions about cosmic rays from
traditional astrophysical sources as well as about exotic cosmic
rays.  In particular, there have been searches for
antiparticles (BESS, CAPRICE, HEAT, IMAX, MASS,
TS93), studies of isotopic composition and clock isotopes
(ISOMAX, SMILI), heavy-element composition (TIGER), element
spectra (BESS, CAPRICE, HEAT and
IMAX ~100 GeV/nucleon; ISOMAX to almost 1 TeV/nucleon; ATIC H and He
to $>10$ TeV; TRACER O to Fe to ~100 TeV total energy), and high-energy
spectra (JACEE, RUNJOB).

\begin{figure}
\includegraphics[width=0.8\textwidth]{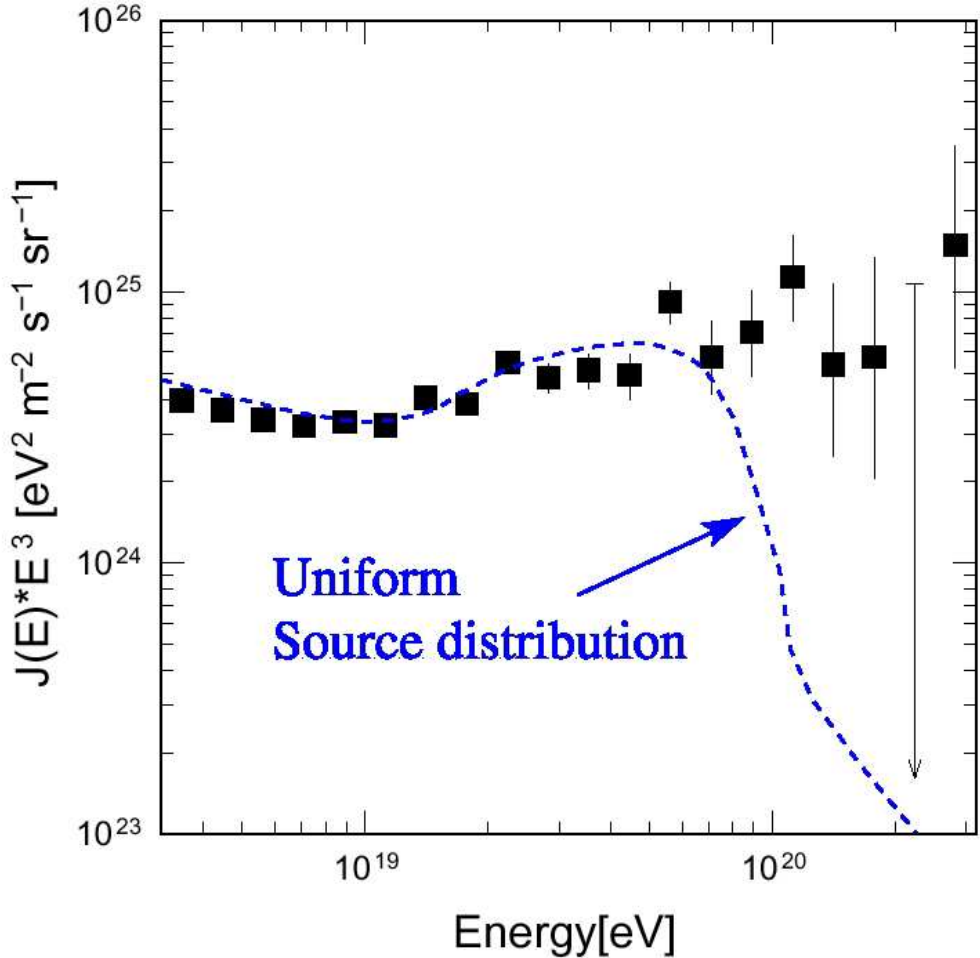}
\caption{Spectrum of ultra-high-energy cosmic
rays measured by AGASA \cite{takada}.}\label{fig-AGASA} 
\end{figure}

\begin{figure}
\includegraphics[width=0.8\textwidth]{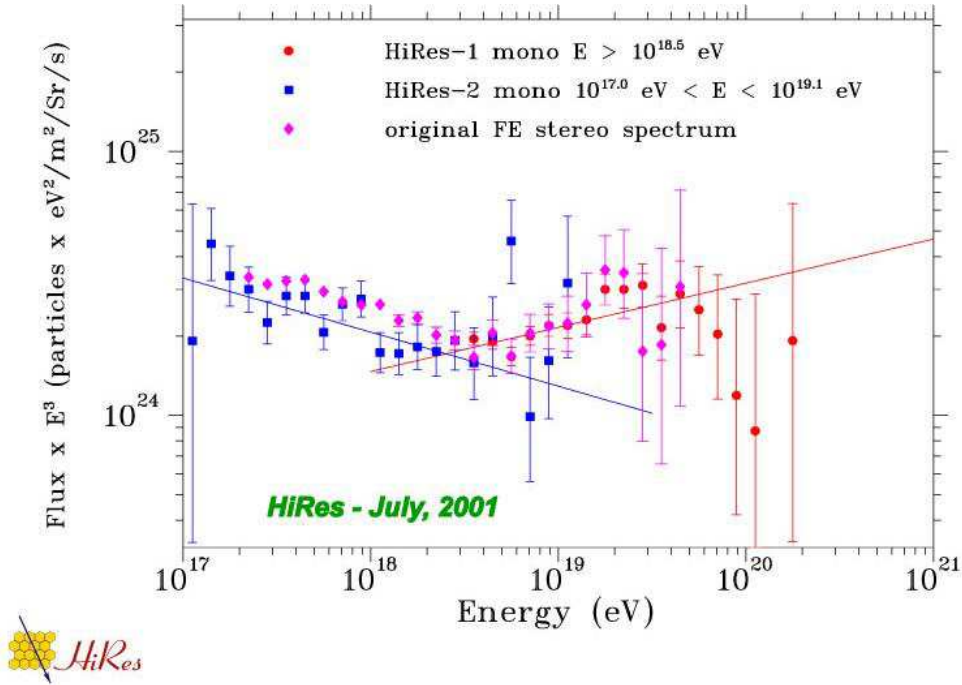}
\caption{\label{fig-hires} Spectrum of ultra-high-energy cosmic
rays from new HiRes data and from early Fly's Eye data.}
\end{figure}

\subsection{Ultra-high-energy cosmic rays}

Of particular interest is the origin and nature of the
ultra-high-energy cosmic rays (UHECRs), those with energies
greater than $3\times10^{19}$ eV, the Greisen-Zatsepin-Kuzmin
(GZK) threshold \cite{Greisen,ZK}.  Such cosmic rays, 
if they are protons, should be
attenuated by photopion production from scattering of CMB
photons and thus have a mean-free path of $\sim30$ Mpc.  Thus,
if the sources of these UHECRs are uniformly distributed
cosmological sources, we would expect to see a drop in
the flux at energies higher than the GZK cutoff, as indicated by
the curve in Figure \ref{fig-AGASA}.  However, observations from
both AGASA \cite{AGASA} (shown in Figure \ref{fig-AGASA})
and HiRes \cite{HiRes} (see Figure \ref{fig-AGASA}) indicate no
such drop; in fact there
may even be an increase in the flux above the GZK threshold
(although there is still some concern about possible
inconsistency between AGASA and HiRes)!  
Another problem is that these cosmic rays (whose trajectories
should be bent negligibly by magnetic fields) do not seem to
come from any obvious accelerators within 30 Mpc.  In fact,
there seem to be no obvious accelerators anywhere within 30 Mpc
of the Milky Way, so even if their trajectories were bent, there
are still no obvious sources.  Moreover, none of these UHECRs
point back to any obvious sources even at larger distances.

There is no single convincing explanation for this mystery,
although there are a number of very intriguing ideas. For example
(1) The apparent excess is due to experimental error and/or
statistical fluctuations.  (2) The primary particles are heavier
nuclei, which might be able to travel farther.  (3) The sources
are nearby astrophysical sources (e.g., intergalactic shocks or
Centaurus A).  Although some of them may have some appeal, all
of these traditional-astrophysics explanations have troubles.
(4) The UHECRs require new physics; e.g., decays of supermassive
cosmological relics or topological defects; some new
non-interacting particles; production of protons from
interactions of ultra-high-energy neutrinos with the 1.9 K
cosmological neutrino background.  The answer could be some
combination of these solutions, or it could be something else
entirely.

This problem has been the focus of much
theoretical \cite{CRReview} and experimental activity, and
major advances can be expected over the next decade.  A summary
of the experimental capabilities is given in Table
I.  Extrapolating the number of events above
$10^{20}$ eV, one would expect $\sim 10$
per year from HiRes, $\sim 100$ per year from Auger 
\cite{Auger} and the telescope array \cite{TelescopeArray}, 
and thousands per year from OWL \cite{OWL}, assuming also
that the spectrum extends beyond $10^{21}$ eV. There is also
EUSO \cite{EUSO}, an ESA mission under study for the space
station, and there may be other novel ideas (e.g., CHICOS
\cite{chicos}) being developed.  Note that the table is
simplistic, and only approximates important effects of duty cycle,
energy thresholds, resolution, and systematics.

\begin{table}
\begin{displaymath}
\begin{tabular}{|c|c|c|c|}
\hline Experiment & Technique & Effective Aperture & Status \\
 & & [$1000\times km^2 sr$] & \\
 \hline \hline AGASA \cite{AGASA} &
Ground Array & $\sim 0.2$ & running \\
 \hline HiRes \cite{HiRes} & Fluorescence & $\sim 1$ & running \\
 \hline Auger \cite{Auger} & Ground Array + Fluor.& 7 & building \\
 & & ($\rightarrow\times 2$) & (proposed) \\
\hline Telescope Array \cite{TelescopeArray} 
& Fluorescence & 8 & under study \\
\hline OWL \cite{OWL} & Fluorescence & $\sim 300$ & under study \\ 
\hline
\end{tabular} \label{table-CR}
\end{displaymath}
\caption{Summary of experiments addressing the mystery of the
highest energy cosmic rays.} \end{table}

\section{Gravitational Radiation}

Almost all current knowledge of the Universe outside the Solar System
derives from observations of electromagnetic radiation (with some
input from neutrinos and cosmic rays).  The advent of
gravitational-wave observatories promises to open a new window on the
astrophysical Universe, likely to lead to a greatly improved
understanding of mysterious phenomena as well as discoveries as yet
completely unanticipated.  A careful survey of gravitational-wave
astrophysics is available in the P4.6 summary report 
\cite{Hughes:2001ch}.

Gravitational radiation is generated by bulk motions of large masses.
The resulting waves induce small tidal forces on any object they pass
by; in interferometric detectors, the most promising current designs,
these tidal forces are detected by measuring the differential
displacement of widely-separated test masses.  The primary problem
facing the prospective gravitational-wave astronomer is the extreme
weakness of the gravitational force.  For a typical source --- e.g., a
coalescing binary neutron star at a distance of several hundred Mpc
--- the strain $h=\delta L/L$ (where $L$ is the distance between the
test masses) is of order $10^{-21}$ or less.  The requisite precision
for ground-based detectors is achieved by means of multiple-pass
interferometry: laser light is first passed through a beam splitter
into the orthogonal Fabry-Perot cavity arms of a Michelson
interferometer, where it makes numerous round trips before being
recombined once more, in such a way that almost all of the light goes
back toward the laser in the absence of a gravitational wave.  A
passing wave disturbs the mirrored test masses, changing the
interference pattern and scattering more light into a photodiode.  Due
to the quadrupole nature of the gravitational waves, the phase shifts
in the orthogonal arms have opposite signs, which can be effectively
detected by interferometric means.  Recycling of the light going back
toward the laser is used to improve the sensitivity.  Dealing with the
various noise sources is one of the greatest challenges of
gravitational-wave astronomy; for ground-based detectors,
detections of coherent signals at
multiple locations will provide strong evidence for the reality of a
passing gravitational wave.

\subsection{Ground-based detectors and their sources}

\begin{figure}
\includegraphics[width=\textwidth]{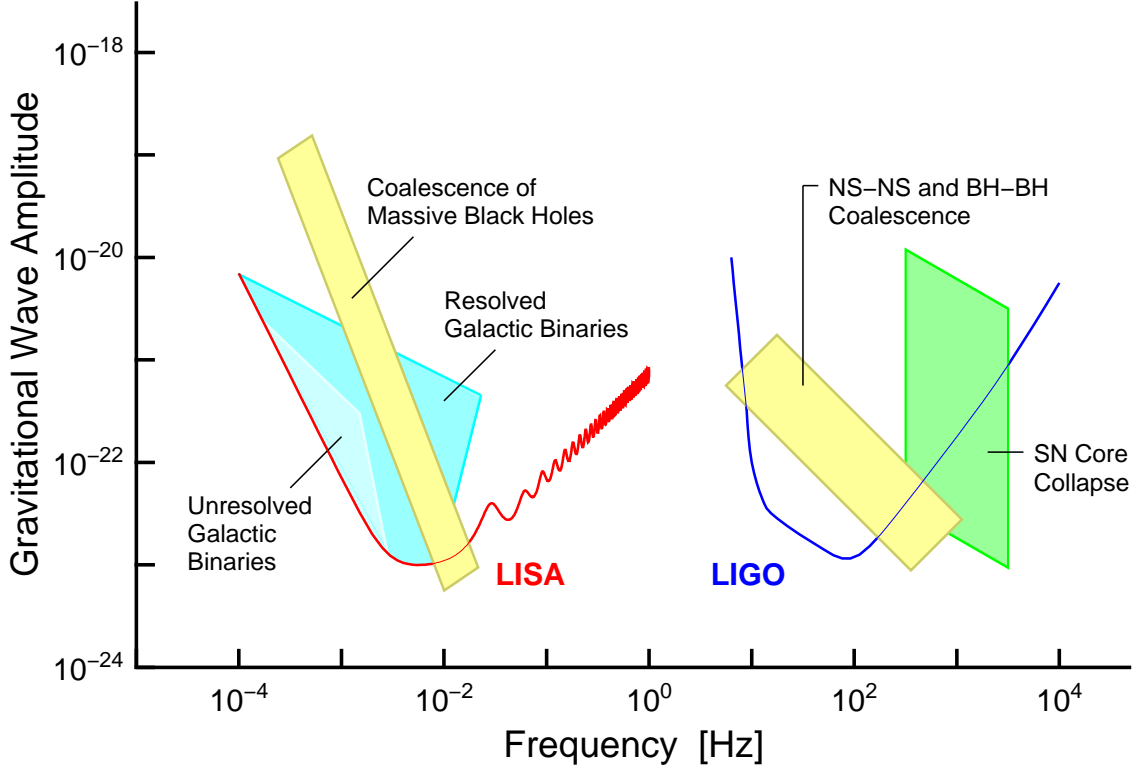}
\caption{\label{fig-ligo-lisa} Sensitivities as a function of
frequency for typical ground-based (LIGO) and space-based
(LISA) gravitational-wave observatories \cite{lisa}.}
\end{figure}

A number of ground-based gravitational-wave observatories are
currently under construction or being commissioned.  In the United
States, these include the two LIGO observatories at Hanford,
Washington and Livingston, Louisiana \cite{ligo}.  Both sites feature
two perpendicular arms of 4 km each; the Hanford site also includes a
pair of 2 km arms sharing the same tunnels as the 4 km arms.  European
facilities include the 3 km interferometer Virgo, in Pisa, Italy
\cite{virgo}, and the 600 meter interferometer GEO600 in Hanover,
Germany \cite{geo600}, both under construction.  Japan has a working
detector, the 300 meter TAMA300 observatory near Tokyo \cite{tama300}.

An important distinction between gravitational and electromagnetic
astronomy is the natural all-sky coverage that gravitational waves
provide.  Determining accurate positional information about a source
of gravitational waves is difficult, and requires information to be
gathered from at least three widely-spaced sites.  A
Southern-hemisphere detector (such as the ACIGA observatory, being
studied in Australia \cite{aciga}) would greatly supplement the
abilities of the existing sites.

Both LIGO observatories have completed their major facility
construction; the 2 km Hanford and 4 km Livingston interferometers
have been installed (and the 4 km Hanford interferometer is well
under way) and are in the process of collecting engineering data,
working toward the beginning of the first science runs.  These
facilities have been purposefully designed to allow for significant
future improvements; major detector upgrades are proposed for 2006.

The limiting noise sources for ground-based detectors depend on the
frequency of the gravitational wave being observed.  Below about 100~Hz
the dominant source is seismic noise, and above 100~Hz it comes from
photon shot noise due to fluctuations in the input laser beam; there
is also a small region in the vicinity of 100~Hz in which thermal
noise in the suspension system for the test masses dominates.  The
``filter'' of observable frequencies runs roughly from 100 to 1000~Hz
for current interferometers, extendable down to 10~Hz in future
upgrades.

In this frequency range, the most promising sources of gravitational
radiation are inspiraling compact binaries, in which both objects are
either black holes or neutron stars \cite{Flanagan:1998sx}.  
For first-generation
interferometers, it is quite possible that no visible sources will be
found, although a black-hole/black-hole binary of 10 solar masses each
at 100~Mpc will be on the edge of detectability.  Future upgrades will
increase sensitivity to new kinds of sources, including rotating
neutron stars \cite{Brady:1998ji}, collapsing stellar cores
\cite{Fryer:2001zw}, and stochastic backgrounds \cite{Maggiore:1999vm}.
(The displacement due to a gravitational wave falls off only as $1/r$,
so that a small increase in sensitivity yields a large increase in the
volume --- second-generation detectors should probe a volume 1000
times bigger than that of first-generation detectors.)  These
instruments will also feature the ability to tune the detectors to
maximize sensitivity in promising frequency ranges.

\subsection{Space-based detectors and their sources}

The only known way to escape low-frequency noise sources on Earth is to
construct detectors in space.  The proposed Laser Interferometer Space
Antenna (LISA) would consist of three satellites, separated by 5
million km, orbiting the Sun at approximately 20 degrees behind the
Earth's position.  A collaboration between NASA and the European Space
Agency, LISA is planned to be launched in 2011
\cite{lisa,Danzmann:1994mm}.

The strain sensitivity of LISA is actually comparable to
that of LIGO, although the frequency ranges are quite different,
with LISA being most sensitive to frequencies around $10^{-2}$~Hz.
In this vicinity, one of the most important sources will be
compact binaries in our galaxy, which actually will provide a noise
background at low frequencies.  Other anticipated astrophysical
sources involve massive (of order $10^6~M_\odot$) black holes
in other galaxies.  LISA will be sensitive to 
a variety of processes, including
the growth and formation of such objects, radiation from smaller
compact objects orbiting around them, and from the coalescence of
two massive black holes.

From the point of view of particle physics, perhaps the most
intriguing potential for LISA is the chance to observe a 
gravitational-wave background from a phase transition in
the early Universe.  If the electroweak phase transition is strongly
first-order (which is disfavored by current models,
but seems to be required in scenarios of electroweak
baryogenesis), the resulting gravitational radiation has a 
frequency today of $10^{-4}$ -- $10^{-3}$~Hz, accessible to LISA
\cite{KKT94,Apreda:2001us}.
Thus, space-based gravitational-wave detectors could provide
crucial information about the electroweak sector.

\subsection{Testing general relativity}

Gravitational-wave observatories offer a direct view into the
behavior of gravity under extreme conditions, especially near
the event horizon of a black hole.  Almost all current tests
of general relativity are carried out in the weak-field regime,
the binary pulsar being the unique exception;
consequently,
observations of gravitational radiation will provide new tests
of classical gravity.  These tests will come both from observing
the behavior of coalescing black holes, and from small black
holes orbiting supermassive ones.  This last test is especially
promising, as it provides a way to map out the curvature of
spacetime using objects which can be successfully theoretically
modeled \cite{Hughes:1999nv}.  
Additionally, we will be testing for the first time
the propagation of gravitational waves across cosmological
distances; comparing this propagation with that of electromagnetic
radiation (does gravity ``travel at the speed of light''?)
will constrain models of extra dimensions as well as
alternative theories of gravity \cite{Caldwell:2001ja}.

\section{Neutrino Astrophysics}

\subsection{Atmospheric and solar neutrinos}

\begin{figure}
\includegraphics[width=\textwidth]{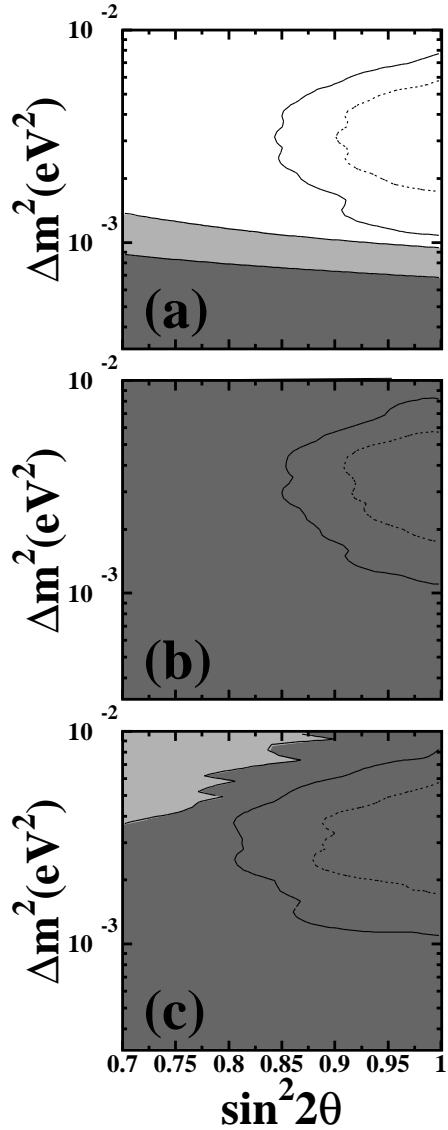}
\caption{\label{fig:nus}  Regions of the mass-mixing parameter
space for mixing into $\tau$ and sterile neutrinos constrained
by atmospheric-neutrino results from super-Kamiokande.  These
results demonstrate that astrophysical particle sources (in this
case, cosmic-ray spallation products) can provide precise and
unique constraints to particle properties.}
\end{figure}

Neutrino astrophysics has recently had a tremendous impact on our
knowledge of neutrino masses and mixing
angles. Recent experiments with atmospheric and solar neutrinos
have started delivering on the promise of using astrophysical sources
to address fundamental particle properties. Since Snowmass96,
the Super-Kamiokande experiment, with its zenith-angle
measurement of the atmospheric neutrino signal, has produced the
clearest evidence to date for a non-trivial neutrino
sector~\cite{superk1999}.  By extending the reach of earlier water
Cerenkov experiments to detection of multi-GeV neutrinos, the SuperK
collaboration has used the angular dependence of muon and electron
neutrinos produced in the atmosphere to obtain a clear signature of
muon neutrinos oscillating into tau neutrinos~\cite{superk2000}
(Fig. \ref{fig:nus}); the data are so precise that there is
even evidence that discriminates between oscillation to tau and to
sterile neutrinos.  This year the SNO collaboration, by combining the
charged-current breakup of the deuteron with elastic scattering on
electrons, provided the first measurement of the electron neutrino
component of the solar-neutrino flux~\cite{sno}. They found that the
solar-neutrino flux is composed of roughly one-third electron
neutrinos and two-thirds other active flavors. By measuring the total
solar-neutrino flux they also confirmed the prediction of the Standard
Solar Model.

What we can infer from the combined SuperK and SNO measurements is
somewhat surprising: it appears that neutrino mixing angles are rather
large, unlike the mixing among quarks. In addition these experiments
provide rather strong constraints for the mixing between first and
second generations, as well as the mixing between second and third
generations. At this time the amount of mixing between the first and
third generations (i.e. the test of unitarity of the neutrino mixing
matrix) and the possible existence of sterile neutrinos that may mix
with the active flavors remain as open questions.  When combined
with results from a variety of laboratory searches for neutrino
masses and mixings, strong constraints to vast regions of
mass-mixing parameter space can be derived (as discussed further
in the P2 sessions on flavor physics).  Further experiments on
atmospheric and solar neutrinos, as well as from other
astrophysical sources such as supernovae and extragalactic
sources, will all help to constrain the properties of the
neutrinos~\cite{bahcall}.

\subsection{Neutrinos elsewhere in the Universe}

Although solar and atmospheric neutrinos have taken center stage
in neutrino astrophysics recently, neutrinos make important
appearances in a variety of other areas in cosmology and
astrophysics.  For many years, concordance between the observed
abundances of light elements and those predicted by BBN provided
the most stringent constraints to the number of light
neutrinos.  Nucleosynthesis and shock reheating in supernovae
also depend on the properties of neutrinos, and can provide
constraints to neutrino masses and mixings.  The CMB and
structure formation may also constrain neutrino properties.

We are also now in the midst of rapid progress in the study of
high-energy astrophysical neutrinos, from the several GeV to PeV
range \cite{gaisser}.  Ocean water or Antarctic ice can provide huge volumes
for neutrino telescopes that detect the Cerenkov radiation
from upward muon induced by energetic neutrinos that interact in
the rock below the detector.  Just in the past two years, the
AMANDA~\cite{amanda} experiment, which operates at the South
Pole, has begun to see a large number of events ($\sim1/day$).
The IceCube~\cite{icecube} experiment, will soon expand on this
success, eventually aiming toward a km$^3$ detector.  Any source
that accelerates high-energy particles should produce
high-energy neutrinos, as high-energy protons (such as those in
Galactic cosmic rays or from extragalactic UHECRs) will interact
with photons or other protons and to produce pions that then
decay to neutrinos.  Thus, possible sources of high-energy
neutrinos (in addition to the numerous atmospheric neutrinos)
include active galactic nuclei and gamma-ray bursts.  Such
neutrino telescopes should also have some sensitivity (at the
$\sim100$ GeV range) to WIMP annihilation from
the Sun or Earth. In addition, there could be even more
interesting surprises since the energy range above a TeV is just
started to be explored. In particular, cosmic-ray protons above
an energy of 10$^{20}$~eV (the ``GZK cutoff'') interact with CMB
photons leading to photoproduction, {\it e.g.} at the $\Delta$
resonances. The resulting neutrino decay products are therefore
a useful probe of the extragalactic processes that would have
produced UHECRs.


\section{The Early Universe and Tests of Fundamental Physics}

Perhaps the most obvious connection between particle physics
and cosmology arises in the study of the early Universe, which
both requires particle physics for a complete description, and
constrains new ideas by requiring that they give rise to a sensible
cosmology.  Here we give a brief overview of this
connection; more details can be found in the P4.8 summary
report~\cite{Albrecht:2001xp}.

\subsection{The Universe as a particle-physics laboratory}

The big-bang model is a framework which, when combined with the
standard model of particle physics, leads to definite predictions for
quantities such as the relic abundances of various particle species.
Comparing these predictions with observations provides crucial
information about the strengths and weaknesses of the standard model.
A canonical example is the asymmetry of baryons over antibaryons,
which Sakharov pointed out can only arise in an initially symmetric
Universe if there are appreciable violations of baryon number, C and
CP symmetries, and thermal equilibrium.  Although in principle all of
these ingredients are to be found in the minimal standard model, in
practice only nonminimal extensions can lead to a quantitatively
acceptable baryon number \cite{Riotto:1998bt}.  Thus, the predominance
of matter over antimatter in the Universe is a strong indication of
the need for new particle physics.

Another such indication is the existence of nonbaryonic dark matter.
Both primordial nucleosynthesis \cite{Schramm:1998vs} and CMB
anisotropies \cite{Pryke:2001yz,boom} are consistent
with a density parameter in baryons $\Omega_b = 0.04 \pm 0.01$,
while dynamical measurements of the matter density yield $\Omega_m
= 0.3 \pm 0.1$, necessitating nonbaryonic matter.  There are no
reasonable candidates for nonbaryonic dark matter in the minimal
standard model.  Massive neutrinos, which are easily accommodated in
minor extensions of the standard model, can contribute to the dark
matter, but their masses are likely to be sufficiently low that they
act as hot dark matter, which is ruled out by studies of large-scale
structure and the CMB.  Instead we must turn to more exotic
possibilities, such as axions or neutralinos (stable superpartners to
ordinary neutral particles) \cite{Ellis:1998gt}.  Direct searches in
terrestrial laboratories for such particles represent a promising
strategy for learning about particle physics in a way that complements
traditional high-energy accelerators; this is discussed in more detail
in Section III.

We have also discussed in Section VI the discovery of dark energy,
a slowly-varying smooth component constituting about $70\%$ of the
critical density.  The simplest candidate for the dark energy is
a cosmological constant $\Lambda$, or vacuum energy, which arises
naturally when quantum field theory is combined with general relativity
\cite{Carroll:2000fy}.
Unfortunately the necessary value of the vacuum energy density is
$\rho_\Lambda \approx 10^{-120}M_{\rm Planck}^4$, far below its
natural value.  This problem is further exacerbated in supersymmetric
theories, where the numerical discrepancy is not as bad (sixty orders
of magnitude if supersymmetry is broken at the weak scale, rather than
120 orders of magnitude), but the reliability of the estimate is much
greater (the vacuum energy is calculable in a given supersymmetric
field theory, rather than simply being an arbitrary parameter).  
In addition, vacuum energy scales rapidly with respect to
matter, $\rho_\Lambda /\rho_{\rm M} \propto a^3$, where $a$ is the
cosmic scale factor.  Since $a$ has changed enormously since the early
Universe (by a factor of order $10^3$ since recombination, and $10^9$
since nucleosynthesis), it is very hard to understand why the vacuum
and matter densities are comparable today.  Attempting to solve this
puzzle has led to numerous proposals for dynamical dark-energy
candidates; to date, none has provided a compelling solution to this
naturalness problem.  An even more dramatic possibility would be to
modify Einstein's general relativity on cosmological scales; but
again, no attractive possibilities have been proposed.

\subsection{Inflation}

The idea that the Universe underwent a period of quasi-exponential
expansion at early times solves a number of cosmological problems,
including the absence of noticeable spatial curvature, the absence of
a significant density of magnetic monopoles from the breaking of a
grand-unified symmetry at high scales, and the apparent violation of
causality in the large-scale homogeneity and isotropy of the Universe
\cite{Liddle:1999mq}. One of the drawbacks of the inflationary 
scenario is that it seems to
inevitably require fine-tuning in the construction of a
particle-physics model responsible for the inflation
\cite{Adams:1991pn}; indeed, despite
a great deal of effort, the search for believable models has not been
a great success.  As particle physics learns more about what lies
beyond the standard model, it may become possible to 
construct a sensible picture of the inflationary era.

Soon after inflation was proposed, it was realized that inflation
could produce a spectrum of nearly scale-free adiabatic perturbations.
It is just such perturbations which seem to be indicated by modern
measurements of CMB anisotropy, as discussed in section V.  Future
measurements of CMB polarization will enable us to disentangle the
effects of cosmological parameters from those of features in the
primordial spectrum, and detection of tensor perturbations
(gravitational waves) will
directly probe a prediction of inflation.  However, it is important to
recognize that exotic models of the inflationary phase (including
multiple fields, non-standard kinetic terms, a time-dependent
gravitational constant, and so on) can yield
perturbations (both scalar and tensor)
with properties that differ in important ways from those of
standard inflation.  Since our knowledge of the physics underlying
inflation is so primitive, we should not be surprised to detect
deviations from the models we consider to be the most simple and
generic; it will be the duty of theorists to continue to refine these
predictions as the experiments continue to improve.

\subsection{Planck-scale physics}

A complete understanding of the early Universe will eventually entail
an understanding of quantum gravity and physics at the Planck scale;
conversely, cosmology provides a unique window onto Planckian physics.
The leading candidate for reconciling quantum mechanics and general
relativity is string theory.  String theory predicts a number of
intriguing features for cosmologists, including supersymmetry,
extended objects (branes), and extra dimensions; however, our current
understanding of the theory is insufficiently developed to enable us
to say anything definite about the early Universe, or cosmological
backgrounds more generally.  Nevertheless, a number of attempts have
been made in this direction; proposals include the idea that the
dynamics of string gases in compactified geometries can explain why
three spatial dimensions are large \cite{Brandenberger:1989aj}, that
the expanding Universe in which we live was preceded by a contracting
phase at early times \cite{Gasperini:1993em}, and that the hot big
bang arose from the collision of initially cold and nearly featureless
three-dimensional branes \cite{Khoury:2001wf}.  While all of these are
speculative and beyond the reach of experimental tests at this time, it
is important to continue to push ideas in this direction, in hopes of
reconciling long standing puzzles of early-Universe cosmology.

Another approach to exploring quantum gravity is to be more
independent of any specific theory, and to analyze possible effects of
generic departures from low-energy physics at the Planck scale.  A
recently popular effort in this direction involves the study of
modified dispersion relations for massless particles, including new
terms that are suppressed by the Planck energy.  While such terms are
clearly negligible in the laboratory, cosmology offers at least two
ways they can be studied: in gradual effects on the propagation of
light as it travels over cosmological distances
\cite{Amelino-Camelia:1998gz}, and in the distortion of modes whose
wavelengths were shorter than the Planck length at early times, and
were subsequently boosted by inflation to give the density
perturbations that have grown into large-scale structure in the
current Universe \cite{Martin:2000xs,Niemeyer:2000eh}.  These two
examples serve to illustrate how cosmological observations may be lead
to direct experimental data that bears on the physics of quantum gravity.

\subsection{Extra dimensions and tests of general relativity}

Much of our speculation about the early Universe relies on an
extrapolation of low-energy physics to the scale of grand unification
or quantum gravity.  While such an extrapolation is perfectly
reasonable, it is not foolproof, and we should be open to the
possibility of dramatically different scenarios.  One example is
provided by brane-world pictures, in which standard-model fields are
confined to a three-dimensional brane embedded in a larger space.
There are two interesting versions of this idea: multiple flat (or
slightly-curved) dimensions of size 1~mm or less (Arkani-Hamed,
Dimopoulos and Dvali, or ADD \cite{Arkani-Hamed:1998nn}), and a single
highly curved extra dimension of arbitrary size (Randall and Sundrum,
or RS \cite{Randall:1999vf}).

Both possibilities come with cosmological implications.  In the RS
model, the conventional Friedmann equation of cosmology can be altered
at high densities, leading to constraints from achieving ``normalcy'' by
big-bang nucleosynthesis.  Modifications of the RS idea have been put
forward in pursuit of a solution to the cosmological-constant problem.
The ADD scenario predicts a plethora of new light particles, on which
cosmology and astrophysics provide stringent bounds.  At the same
time, the dynamics of the extra dimensions can play an important role
in processes such as inflation and baryogenesis.

These scenarios can lead in principle to deviations from conventional
gravity on short (but macroscopic) length scales.  This has led to
renewed interest in laboratory tests of Newtonian gravity.  In
particular, the ADD scenario with $n$ extra dimensions predicts a
changeover from a $1/r^2$ force law to a $1/r^{2+n}$ law at distances
smaller than the size of the extra dimensions; recent searches for
this effect have limited this size to less than 0.1~mm
\cite{Hoyle:2000cv}, and
future experiments will continue to improve this limit.

The possibilities of extra dimensions and new light scalar fields
are strongly suggested by string theory, and bring home the 
importance of a wide-ranging program to search for deviations
from general relativity.  Light scalar fields will generically
couple to ordinary matter with Planck-suppressed interactions,
which can lead to fifth-force effects and apparent violations of
the Equivalence Principle.  Laboratory experiments have placed
stringent bounds on such phenomena, and proposed satellite and
radar-ranging tests would bring a significant improvement.  
Coupled with observations from gravitational-wave observatories,
observations of binary pulsars, and new satellite and laboratory
tests, these experiments promise a new era of precision
knowledge of the behavior of gravity.

\section{Summary and conclusions}

Particle astrophysics and cosmology present a number of exciting
theoretical and experimental challenges along a broad front.
There are a variety of mysterious astrophysical and cosmological
phenomena that almost certainly will require new physics to be
understood. Here we have surveyed much of the activity in this
field, focusing in particular on areas with recent
breakthroughs, and presented a broad overview of some of the
most exciting questions that may be addressed experimentally in
the foreseeable future.  More detailed discussions of all of the
topics we have covered can be found in the topical subgroup
reports in this Snowmass proceedings.

\begin{acknowledgments}
We thank the several hundred Snowmass attendees who participated
in the P4 sessions, as well as in the closely aligned E6
sessions.  We thank in particular our topical sub-group
organizers, as well as N. Mavalvala.
MK was supported by NSF AST-0096023, NASA NAG5-8506, and DoE
DE-FG03-92-ER40701 and DE-FG03-88-ER40397.  SC was supported by
DOE DE-FG02-90ER-40560,  the Alfred P. Sloan Foundation, and the
David and Lucile Packard Foundation. DSA acknowledges the
support of the NSF CAREER program Grant No. PHY-9722414 and the
Center for Particle Astrophysics, an NSF Science and Technology
Center operated by the University of California, Berkeley, under
Co-operative Agreement No. AST-91-20005.

\end{acknowledgments}

\bibliography{p4_convenors}

\end{document}